\documentclass{emulateapj}

\usepackage{float} 
\usepackage{amsmath}
\usepackage{epsfig,floatflt}
\usepackage{subfigure}
\usepackage{rotating}

\begin{document}

\title{The $N$-point correlation functions of the first-year
  \emph{Wilkinson Microwave Anisotropy Probe} sky maps}

\author{H.\ K.\ Eriksen\altaffilmark{1,2,3}} \affil{Institute of
Theoretical Astrophysics, University of Oslo, P.O.\ Box 1029 Blindern,
\\ N-0315 Oslo, Norway}
\altaffiltext{1}{Also at Centre of Mathematics for Applications,
University of Oslo, P.O.\ Box 1053 Blindern, N-0316 Oslo}
\altaffiltext{2}{Also at Jet Propulsion Laboratory, M/S 169/327, 4800
  Oak Grove Drive, Pasadena CA 91109} 
\altaffiltext{3}{Also at California Institute of Technology, Pasadena, CA
91125}
\email{h.k.k.eriksen@astro.uio.no}

\author{A.\ J.\ Banday}
\affil{Max-Planck-Institut f\"ur Astrophysik, Karl-Schwarzschild-Str.\
1, Postfach 1317,\\D-85741 Garching bei M\"unchen, Germany} 
\email{banday@MPA-Garching.MPG.DE}

\author{K.\ M.\ G\'orski\altaffilmark{3}} 
\affil{Jet Propulsion Laboratory, M/S 169/327, 4800 Oak Grove Drive, 
Pasadena CA 91109\\ Warsaw University Observatory, Aleje Ujazdowskie
4, 00-478 Warszawa, Poland}

\email{Krzysztof.M.Gorski@jpl.nasa.gov}

\and

\author{P.\ B.\ Lilje\altaffilmark{1}} \affil{Institute of Theoretical
Astrophysics, University of Oslo, P.O.\ Box 1029 Blindern, \\N-0315
Oslo, Norway}

\email{per.lilje@astro.uio.no}


\begin{abstract}
We compute the two-, three- and four-point correlation functions from
the \emph{Wilkinson Microwave Anisotropy Probe} (\emph{WMAP})
first-year data, and compare these to a Monte Carlo ensemble of 5000
realizations, based on the best-fit \emph{WMAP} running-index spectrum
of Gaussian fluctuations. The analysis is carried out in three steps,
covering small ($<72\arcmin$), intermediate ($<5^{\circ}$) and large
scales (up to $180^{\circ}$). On the largest scales our results are
consistent with the previously reported hemisphere power asymmetries:
the northern ecliptic hemisphere is practically devoid of large scale
fluctuations, while the southern hemisphere show relatively strong
fluctuations. We also detect excess correlations in $W$-band
difference maps as compared to the detailed noise simulations produced
by the \emph{WMAP} team, possibly indicative of unknown
systematics. While unlikely to affect any temperature based results,
this effect could potentially be important for the upcoming
polarization data. On intermediate angular scales we find hints of a
similar anisotropic distribution of power as seen on the very largest
scales, but not to the same extent. In general the model is accepted
on these scales. Finally, the same is also true on the smallest scales
probed in this paper.
\end{abstract}


\maketitle

\section{Introduction}

In recent months, a large number of analyses focusing on
non-Gaussianity in the \emph{Wilkinson Microwave Anisotropy Probe}
(\emph{WMAP}; Bennett et al.\ 2003a) data have claimed significant
detections of non-Gaussian features \citep{copi:2004, de
Oliveira-Costa:2004, eriksen:2004a, eriksen:2004b, hansen:2004a,
hansen:2004b, larson:2004, mcewen:2004, park:2004, vielva:2004}. If
any one of these detections can be shown to be of cosmological origin,
currently accepted models based on Gaussianity and isotropy would have
to be revised. Gaining proper understanding of their nature is
therefore essential for further progress.

Sources of non-Gaussian (or anisotropic) signal may be categorized
into three general classes. First, most non-cosmological foregrounds
are highly non-Gaussian, and are all likely to introduce a
non-Gaussian signal into the maps to some extent. In fact, unless some
particular detection is explicitly demonstrated to be
frequency-independent, it must usually be assumed to be
foreground-induced. Second, systematics may introduce non-Gaussian
signals into the data. An example of this is correlated noise, that
results in stripes along the scanning path of the experiment. Finally,
the most intriguing possibility is that the primordial density field
itself could be non-Gaussian, e.g., through the existence of
topological defects or non-equilibrium inflation.

In the current paper, we subject the \emph{WMAP} data to an analysis
based on real-space $N$-point correlation functions. While
harmonic-space methods often are preferred over real-space methods for
studying primordial fluctuations, real-space methods may have an
advantage with respect to systematics and foregrounds, since such
effects are usually localized in real space. It is therefore important
to analyze the data in both spaces in order to highlight different
features. For instance, by considering difference maps between
independent differencing assemblies (abbreviated DA; see Hinshaw et
al.\ [2003] for details on the terminology), that ideally should
contain no CMB signal, we detect excess correlations in the data that
are not accounted for in detailed simulations of the \emph{WMAP}
pipeline, and by partitioning the sky into small regions, we find
hints of residual foregrounds near the galactic plane.

The algorithms used in this paper were developed by
\citet{eriksen:2004c}, and applied to the first-year \emph{WMAP} data
by \citet{eriksen:2004a}. Other $N$-point correlation function
analyses of the first-year \emph{WMAP} data include those presented by
\citet{gaztanaga:2003}, \citet{gaztanaga:2003b}, and \citet{land:2004}.

\section{Definitions}
\label{sec:definitions}

The statistics of interest in this paper are the $N$-point correlation
functions (here restricted to two-, three- and four-point functions),
and we measure these functions both for the observed data and for an
ensemble of simulated realizations with controlled properties. A
$\chi^2$ statistic is then employed to quantitatively measure the
agreement between the data and the model.

An $N$-point correlation function is by definition the average product
of $N$ temperatures, measured in a fixed relative orientation on the
sky,
\begin{equation}
C_{N}(\theta_{1}, \ldots, \theta_{2N-3}) = \biggl<\Delta
T(\hat{n}_{1}) \cdots
\Delta T(\hat{n}_{N}) \biggr>,
\label{eq:npoint_def}
\end{equation}
where the unit vectors $\hat{n}_1, \ldots, \hat{n}_{N}$ span an
$N$-point polygon on the sky. By assuming statistical isotropy, the
$N$-point functions are only functions of the shape and size of the
$N$-point polygon, and not on its particular position or orientation
on the sky. Hence, the smallest number of parameters that
uniquely determines the shape and size of the $N$-point polygon is
$2N-3$.

The $N$-point correlation functions are estimated by simple product
averages,
\begin{equation}
C_{N}(\theta_{1}, \ldots, \theta_{2N-3}) = \frac{\sum_i w_1^i \cdots
  w_N^i \cdot T_1^i \cdots T_N^i}{\sum_i w_1^i \cdots w_N^i },
\end{equation}
where the sums are taken over all sets of $N$ pixels fulfilling the
geometric requirements set by $\theta_1, \ldots, \theta_{2N-3}$. The
pixel weights, $w_i$, may be independently chosen for each pixel in
order to reduce, e.g., noise or border effects. Here they represent
masking, by being set to 1 for included pixels and to 0 for excluded
pixels.

The main difficulty with computing $N$-point functions is their
computational scaling. The number of independent pixel combinations
scales as $\mathcal{O}(N_{\textrm{pix}}^N)$, and for each combination
of $N$ pixels, $2N-3$ angular distances must be computed to uniquely
determine the properties of the corresponding polygon. Computing the
full $N$-point function for $N>2$ and $N_{\textrm{pix}} \gtrsim 10^5$
is therefore computationally challenging.

However, it is not necessary to include all possible $N$-point
configurations in order to produce interesting results. E.g.,
one may focus only on small angular scales, or on
configurations with some special symmetry properties. By using the
methods described by \citet{eriksen:2004c}, the computational expense
then becomes tractable, since no CPU time is spent on excluded
configurations. In this paper several such subsets are computed,
covering three distinct ranges of scales, namely small (up to
$1\fdg 2$), intermediate (up to $5^{\circ}$) and large scales (the
full range between $0^{\circ}$ and $180^{\circ}$).

\subsection{The $\chi^2$ statistic}
\label{sec:chisquare}

In this paper, a simple $\chi^2$ test is chosen to quantify the degree
of agreement between the simulations and the observations, where
$\chi^2$  as usual is defined by
\begin{equation}
\chi^2 = \sum_{i,j=1}^{N_{\textrm{bin}}} (C_N(i) - \bigl<C_N(i)\bigr>)M_{ij}^{-1}(C_N(j) -
\bigl<C_N(j)\bigr>).
\label{eq:chisquare}
\end{equation}
Here $C_N(i)$ is the $N$-point correlation function for
configuration\footnote{The terms 'configuration' and 'bin' are used
interchangeably in this paper.} number $i$, $\bigl<C_N(i)\bigr>$ is the
corresponding average from the Monte Carlo ensemble, and
\begin{equation}
M_{ij} = \frac{1}{N_{\textrm{sim}}}\sum_{k=1}^{N_{\textrm{sim}}}
(C^{(k)}_N(i) - \bigl<C_N(i)\bigr>)(C^{(k)}_N(j) - \bigl<C_N(j)\bigr>)
\end{equation}
is the covariance matrix.

This statistic is optimized for studying Gaussian distributed
data. Unfortunately, the $N$-point correlation functions (and in
particular even-ordered ones) are generally strongly non-Gaussian (and
asymmetrically) distributed, and this leads to an uneven weighting of
the two tails by the $\chi^2$ statistic. In order to remedy this
weakness, the empirical distribution of each
configuration is transformed by the relation \citep{eriksen:2004a}
\begin{equation}
\frac{\textrm{Rank of observed map}}{\textrm{Total number of maps}+1}
= \frac{1}{\sqrt{2\pi}} \int_{-\infty}^{s}
e^{-\frac{1}{2} t^2} dt.
\label{eq:gaussianize2}
\end{equation}
The numerator of the left hand side is the number of realizations with
lower value than the current map, and the denominator is the total
number of realizations plus 1. The addition of 1 is necessary to
obtain symmetric values of $s$ around 0, and to avoid that the
realization with the lowest value is assigned an infinite confidence
level. Note that if the data were in fact Gaussian distributed,
equation \ref{eq:gaussianize2} is an identity operation in the limit
of an infinite number of simulations. The $\chi^2$ statistic is then
computed from the transformed data, rather than from the original
correlation functions.

The quoted significance level is given in terms of the fraction of
simulations with a lower $\chi^2$ value than the observed map. Thus, a
value more extreme than either 0.025 or 0.975 indicates that the model
is rejected at the $2\, \sigma$ level.

In order to eliminate any procedural difference between the simulations
and the observed maps, we include the observed map itself
in the estimation of the covariance matrix. While this should have no
impact on the result if the covariance matrix is properly converged,
it is a very useful safe-guard against such issues.

A singular value decomposition (SVD) is used to compute the
inverse covariance matrix, and conservatively all modes with a
condition number smaller than $10^{-6}$ are set to zero. However, this limit
is only reached in the small-scale analysis, in which different
neighboring configurations are very strongly correlated, and the
covariance matrix converges more slowly than for the intermediate- and
large-scale functions.

Finally, the four-point correlation function is treated differently
than the two- and three-point functions, in that its power
spectrum dependence is reduced by utilizing the following relationship: if a
random field is Gaussian, then the ensemble average of the four-point
function may be written in terms of the two-point function (see, e.g.,
Adler 1981),
\begin{align}
\bigl<T_1 T_2 T_3 T_4\bigr> = & \bigl<T_1 T_2 \bigr>\bigl<T_3 T_4 \bigr>
+ \bigl<T_1 T_3 \bigr>\bigl<T_2 T_4 \bigr> \\
& \quad + \bigl<T_1 T_4
\bigr>\bigl<T_2 T_3 \bigr>.
\end{align}
We therefore subtract the quantity on the right-hand side from the
observed four-point function, to obtain a reduced four-point
function. In what follows, all $\chi^2$ results for the four-point
function refer to this reduced function.

\section{Preparation of the data}
\label{sec:prep}

The first-year \emph{WMAP} data may be downloaded from
LAMBDA\footnote{http://lambda.gsfc.nasa.gov}. Most of the analyses
described in the following sections are carried out for both the raw
maps and the template-corrected versions \citep{bennett:2003b}.

\begin{figure}
\mbox{\epsfig{file=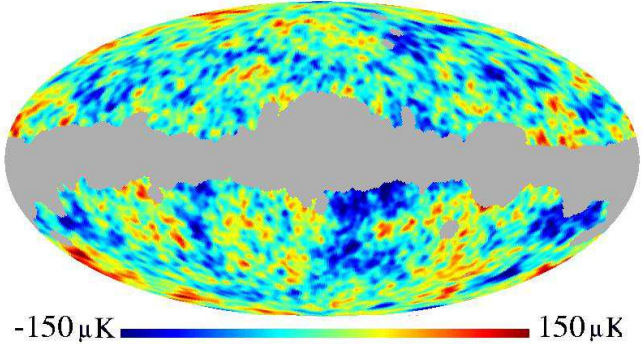,width=\linewidth,clip=}}
\caption{The low-resolution co-added \emph{WMAP} map is made by
  smoothing each of the eight cosmologically interesting bands to a
  common $\textrm{FWHM}=140'$ Gaussian beam, and subsequently
  co-adding these using inverse-noise weights. Finally, best-fit
  monopole, dipole, and quadrupole moments were removed from the
  high-latitude region.}
\label{fig:wmap_nside0064}
\end{figure}

We define our model for the simulations as the sum of a CMB component
and a noise component. The signal component is based on the best-fit
\emph{WMAP} power spectrum with a running index, including multipole
components with $\ell=2,\ldots,1024$, filtered through the
HEALPix\footnote{http://www.eso.org/science/healpix} (G\'orski, Hivon,
\& Wandelt 1999) pixel window functions and channel-dependent beam
windows. While there is some controversy about the evidence for a
running index, we have found that this spectrum provides a better
fit to the data at the very low-$\ell$ range of the spectrum, and
therefore a better fit in terms of $N$-point correlation functions
that are sensitive to large-scale structures. However, this
difference is only noticeable for the two-point function; the
three-point and reduced four-point functions are only mildly dependent
on the assumed power spectrum, thus our results should be independent
thereof. Finally, the $a_{lm}$'s are assumed to be Gaussian.

The noise is assumed to be uncorrelated and Gaussian, with rms levels
given for each pixel of each channel by the \emph{WMAP} team
\citep{bennett:2003a}. This noise is added pixel by pixel and channel
by channel to the CMB signal realizations. 

We study both individual frequency maps and a co-added version that
includes all eight bands. The frequency maps are generated by straight
averaging over bands using equal weights, whereas the co-added map is
weighted with inverse noise variance weights \citep{hinshaw:2003a}.

The analysis is carried out in two steps: First we study the
large-scale fluctuations on the full sky\footnote{Whenever we refer to
a `full-sky' analysis, we mean that data from both hemispheres are
included, except for those pixels excised to avoid contamination from
the galactic plane and point-sources, where appropriate.} by degrading
the maps from $N_{\textrm{side}} = 512$ to $N_{\textrm{side}}=64$,
$N_{\textrm{side}}$ being the HEALPix resolution parameter (G\'orski
et al.\ 1999). We then study the small and intermediate scales by
partitioning the full-resolution sky into disks of $10^{\circ}$ radius
(in two different configurations), and compute the correlation
functions on each disk separately. Full-sky functions for these scales
are estimated by averaging over all disks.

The degradation process may be written on the following algorithmic
form:
\begin{enumerate}
\item Compute the spherical harmonic components, $a_{\ell m}$ from the
  full resolution $N_{\textrm{side}} = 512$ map.
\item Deconvolve with the original \emph{WMAP} beam and pixel windows
  (i.e., multiplication in harmonic space).
\item Convolve with a $140\arcmin$ FWHM Gaussian beam and
  $N_{\textrm{side}} = 64$ pixel windows.  
\item Compute the $N_{\textrm{side}} = 64$ map using the filtered
  $a_{\ell m}$'s.  
\end{enumerate}
This process is carried out for each channel separately before any
co-addition is done. The downgraded, co-added \emph{WMAP} map is shown
in Figure \ref{fig:wmap_nside0064}.

Since all structures in the high-resolution maps are smoothed out in
the degrading process, the foreground exclusion mask must also be
extended correspondingly. This is done by setting all excluded pixels
in the original mask to 0 and all included pixels to 1, then convolving
this map with a Gaussian beam of the desired FWHM, and finally excluding
all pixels with a value smaller than 0.99 in the smoothed mask.

This degraded mask will only be used on smoothed, low-resolution maps,
and the Kp0 mask is therefore used with point sources \emph{not}
excluded as our base mask. This could in principle introduce a
non-Gaussian signal into our maps, but in practice point sources
contribute negligible power at scales larger than a few degrees
\citep{hinshaw:2003a}.

\begin{figure*}

\mbox{\epsfig{file=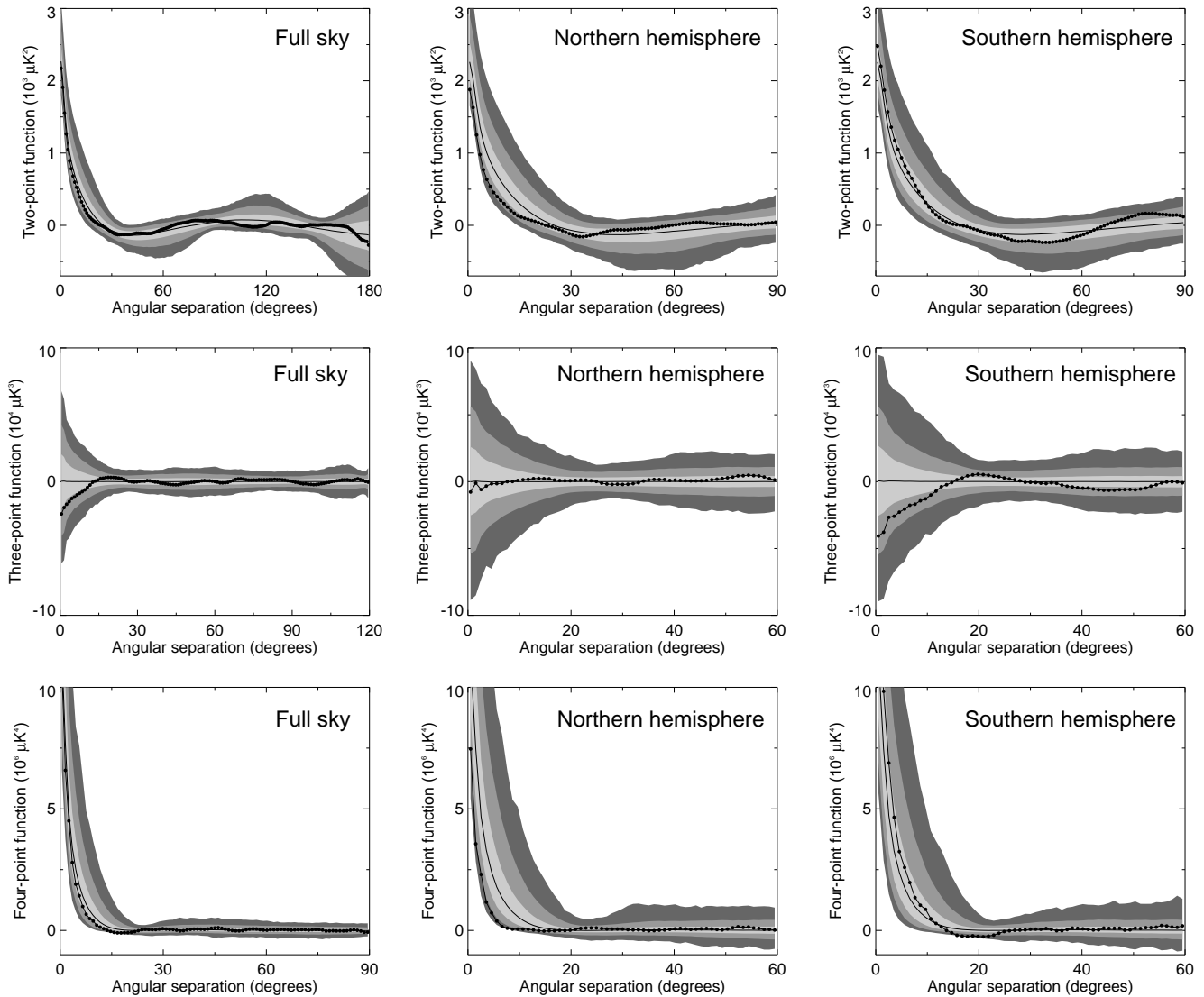,width=\linewidth,clip=}}

\caption{The large-scale, low-resolution correlation functions
  computed from from the co-added \emph{WMAP} map. The solid dots show the
  results from the observed data, the solid line and the gray bands
  show the median and the 1, 2, and $3\, \sigma$ confidence regions,
  respectively, computed from 5000 simulations. The full-sky, northern,
  and southern galactic hemisphere results are plotted in left, middle,
  and right columns, while rows show the two-point, the equilateral
  three-point, and the rhombic four-point functions. Note in
  particular the extremely featureless correlation functions computed
  on the northern hemisphere, indicating little large-scale structure
  in this region. Similar plots for the ecliptic hemispheres are
  shown by \citet{eriksen:2004a}.}
\label{fig:corrfuncs_large}

\end{figure*}

Finally, for the low-resolution analysis, we remove the monopole,
dipole, and quadrupole modes from each map separately, with parameters
computed from the high-latitude regions of the sky only (defined by
the extended Kp0 mask). The reason for removing the quadrupole is that
this particular mode may have an anomalously low value
\citep{bennett:2003a}, but is certainly contaminated by residual
foregrounds after template subtraction
\citep{slosar:2004,hansen:2004b,eriksen:2004d}. Since real-space
correlation function are inherently more sensitive to the low-$\ell$
modes, this well-known effect could mask other interesting features.

In the case of the small- and intermediate-scale analyses, we estimate
the correlation functions on independent disks of $10^{\circ}$
radius. In order to reduce the correlation between neighboring disks,
we therefore choose to remove all multipoles\footnote{The particular
$\ell_{\textrm{max}}=18$ was chosen to correspond roughly to the disk
radius of $10^{\circ}$.} with $0 \le \ell \le 18$, by generalizing the
usual method of removing the low-$\ell$ components to higher
multipoles. 

\section{Large-scale analysis}
\label{sec:large_scales}

The first analysis focuses on the very largest scales, by computing
the $N$-point correlation functions from degraded maps, as described
above. The functions are uniformly binned with $1^{\circ}$ bin size,
and the two-point function is computed over the full range between 0
and $180^{\circ}$. For the higher-order functions we follow
\citet{eriksen:2002} and compute the pseudo-collapsed and the
equilateral three-point functions, and the 1+3-point and the rhombic
four-point functions. 

\begin{figure*}
\epsscale{1.0}
\mbox{\epsfig{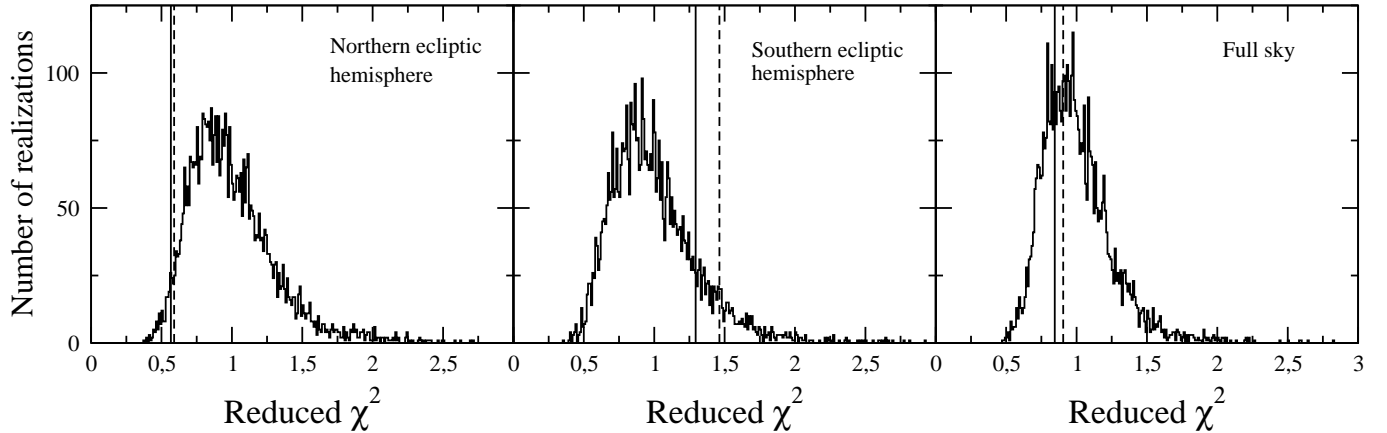}}
\caption{Distributions of the $\chi^2$ values computed from the
  ensemble for the full sky three-point function; the plots show the
  results for the northern ecliptic hemisphere (left), the southern
  ecliptic hemisphere (middle), and the full sky (right). The value
  corresponding to the foreground corrected co-added map is marked
  with a solid line, while the raw co-added map is denoted with a
  dashed line.}
\label{fig:chi_squares}
\end{figure*}

The definition of ``pseudo-collapsed'' is slightly modified compared
to the one described by \citet{eriksen:2002}. In this paper
``pseudo-collapsed'' indicates that the length of the collapsed edge
falls within the second bin, and not that only neighboring pixels are
included \citep{gaztanaga:2003,eriksen:2004a}. This modification
eliminates the need for treating each configuration as a special case,
and is thus purely implementation-ally motivated.

The results from these measurements are shown in Figure
\ref{fig:corrfuncs_large} for the co-added map, for a few selected
functions. A complete summary of the large-scale measurements are
given in Table \ref{tab:chisq_large} for both individual channels and
for the co-added map, and for two different masks. 

Considering first the full-sky two-point function, we see that this
function demonstrates an almost complete lack of structure above
$60^{\circ}$, and its overall shape is very flat as pointed out by
several authors (e.g., Bennett et al.\ 2003a). However, here it is
important to remember that the quadrupole was removed prior to the
computation of the correlation functions, and therefore the two-point
function does not appear quite as anomalous as that seen in many other
plots\footnote{Although the two-point correlation function is the
Legendre transform of the power spectrum, it does not necessarily
follow that the observed two-point function agrees with an ensemble
average based on a power spectrum fitted to the data: the best-fit
power spectrum is largely determined by the small scale information
(high $\ell$'s) in the data, whereas the two-point function is very
sensitive to the largest scales (low $\ell$'s). The two functions thus
provide complementary pictures of the data, highlighting different
features.}.

\begin{deluxetable*}{lcccccccc}
\tablewidth{0pt}
\tabletypesize{\small}
\tablecaption{Large scale $N$-point correlation function $\chi^2$
and $S$-statistic results\label{tab:chisq_large}} 
\tablecomments{Results from $\chi^2$ tests of the large-scale correlation
functions. The numbers indicate the fraction of simulations
with a $\chi^2$ value \emph{lower} than for the respective \emph{WMAP}
map.}
\tablecolumns{9}
\tablehead{ & \multicolumn{2}{c}{$Q$-band} & \multicolumn{2}{c}{$V$-band} &
  \multicolumn{2}{c}{$W$-band} & \multicolumn{2}{c}{Co-added} \\
Region & Kp0 & $|b|>30^{\circ}$ & Kp0 & $|b|>30^{\circ}$ & Kp0 &
$|b|>30^{\circ}$ & Kp0 & $|b|>30^{\circ}$}
\startdata

\cutinhead{Two-point function; $\chi^2$ statistic}
Full sky	                & 0.725 & 0.558 & 0.491 & 0.606 &
0.519 & 0.429 & 0.574 & 0.451 \\[2mm]
Northern galactic	        & 0.495 & 0.605 & 0.721 & 0.732 & 0.682 &
0.574 & 0.772 & 0.578 \\
Southern galactic	        & 0.903 & 0.861 & 0.792 & 0.865 & 0.740 &
0.629 & 0.865 & 0.803 \\[2mm]
Northern ecliptic	        & 0.439 & 0.508 & 0.608 & 0.582 & 0.218 &
0.536 & 0.538 & 0.469 \\
Southern ecliptic	        & 0.272 & 0.729 & 0.215 & 0.676 & 0.125 &
0.572 & 0.216 & 0.598 \\

\cutinhead{Two-point function; $S$-statistic results}
Full sky	                & 0.094 & 0.102 & 0.084 & 0.121 & 0.068 &
0.131 & 0.085 & 0.107 \\[2mm] 
Northern galactic	        & 0.026 & 0.439 & 0.019 & 0.379 & 0.036 & 
0.417 & 0.022 & 0.419 \\
Southern galactic	        & 0.739 & 0.433 & 0.776 & 0.555 & 0.726 &
0.567 & 0.749 & 0.488 \\
Ratio of $S$-values	        & 0.033 & 0.239 & 0.025 & 0.168 & 0.034 & 
0.203 & 0.029 & 0.214\\[2mm]  
Northern ecliptic	        & 0.016 & 0.015 & 0.017 & 0.012 & 0.012 &
0.013 & 0.015 & 0.014 \\
Southern ecliptic	        & 0.745 & 0.693 & 0.791 & 0.784 & 0.743 &
0.794 & 0.759 & 0.735 \\
Ratio of $S$-values	        & 0.045 & 0.075 & 0.038 & 0.052 & 0.048 & 
0.050 & 0.044 & 0.063\\ 

\cutinhead{Three-point function; $\chi^2$ statistic}
Full sky	                & 0.314 & 0.711 & 0.267 & 0.636 &
0.154 & 0.616 & 0.284 & 0.641 \\[2mm]
Northern galactic	        & 0.041 & 0.051 & 0.036 & 0.050 & 0.034 &
0.061 & 0.031 & 0.060 \\
Southern galactic	        & 0.825 & 0.796 & 0.819 & 0.847 & 0.819 &
0.774 & 0.822 & 0.821 \\
Ratio of $\chi^2$'s	        & 0.030 & 0.046 & 0.027 & 0.033 &
0.027 & 0.057 & 0.026 & 0.044 \\[2mm]
Northern ecliptic	        & 0.047 & 0.046 & 0.023 & 0.033 & 0.014 &
0.038 & 0.034 & 0.041 \\
Southern ecliptic	        & 0.831 & 0.792 & 0.861 & 0.820 & 0.871 &
0.829 & 0.840 & 0.793 \\
Ratio of $\chi^2$'s	        & 0.031 & 0.040 & 0.014 & 0.031 &
0.012 & 0.031 & 0.023 & 0.039 \\

\cutinhead{Four-point function; $\chi^2$ statistic}
Full sky	                & 0.484 & 0.518 & 0.491 & 0.480 &
0.474 & 0.472 & 0.468 & 0.508 \\[2mm]
Northern galactic	        & 0.089 & 0.073 & 0.069 & 0.053 & 0.086 &
0.051 & 0.077 & 0.061 \\
Southern galactic	        & 0.884 & 0.920 & 0.905 & 0.930 & 0.878 &
0.914 & 0.888 & 0.923 \\
Ratio of $\chi^2$'s	        & 0.030 & 0.022 & 0.020 & 0.014 & 0.031 &  
0.017 & 0.025 & 0.018 \\[2mm]
Northern ecliptic	        & 0.070 & 0.020 & 0.054 & 0.010 & 0.050 &
0.012 & 0.058 & 0.014 \\
Southern ecliptic	        & 0.852 & 0.927 & 0.873 & 0.942 & 0.846 &
0.931 & 0.857 & 0.932 \\
Ratio of $\chi^2$'s	        & 0.030 & 0.004 & 0.021 & 0.001 &
0.025 & 0.002 & 0.025 & 0.002 

\enddata
\end{deluxetable*}

Next, the full-sky three-point function shows similar tendencies, as
it lies inside the $1\,\sigma$ confidence region almost over the full
range of scales. Finally, the four-point function is quite low at
small angles, and very close to zero at large angles. Thus, all three
full-sky correlation functions point toward the same conclusion --
there is little large-scale power in the \emph{WMAP} data.

Several analyses have presented evidence for a significant asymmetry
between the northern and southern ecliptic (and galactic) hemispheres
\citep{eriksen:2004a,eriksen:2004b,hansen:2004a,hansen:2004b,park:2004},
and therefore we choose to estimate the various functions from these
regions separately. Similar patterns are indeed found also in these
cases: the northern hemisphere correlation functions all show a
striking lack of fluctuations, whereas the southern hemisphere
functions show good agreement with the confidence bands computed from
the Gaussian simulations. This difference translates into a clear
difference of $\chi^2$ numbers, as seen in Table
\ref{tab:chisq_large}. The northern hemisphere results for the
higher-order functions all lie in the bottom few percent range, while
the corresponding numbers for the southern hemisphere are generally
higher than 80\%.

The $\chi^2$ statistic may actually serve as a general measure of the
overall fluctuation level of the higher-order functions, since they
both have vanishing mean (we use the reduced, not the complete,
four-point function in these analyses; for the same reason, this does
not work for the two-point function). One possible statistic for the
degree of power asymmetry between two complimentary hemisphere is
therefore simply the ratio of the two individual $\chi^2$'s. This
quantity is computed for both the simulations and the observed data,
and the fraction of simulations with a smaller ratio is listed in the
third row of each section in Table \ref{tab:chisq_large}. 

We see that this ratio is extreme at the few percent level for the
three-point function, and at less than one percent for the four-point
function, for the ecliptic hemispheres. Further, it is not
particularly sensitive to frequency or galactic cut. In fact, the
numbers are slightly stronger for the conservative $|b|>30^{\circ}$
mask than for the Kp0 mask in four-point function case. Both these
results argue strongly against a foreground based explanation.

In order to quantify the two-point function asymmetry, we adopt a
slightly modified version of the $S$-statistic, as defined by
\citet{spergel:2003}
\begin{equation}
S = \int [C_2(\theta)]^2 \textrm{d}\cos\theta.
\end{equation}
Note that we choose to include the full range of available angles,
while \citet{spergel:2003} chose to exclude angles smaller than
$60^{\circ}$. Excluding the smaller angles does increase the nominal
significance of this statistic when applied to the \emph{WMAP} data,
but it also makes the interpretation of the final results less clear,
since the cut-off scale is arbitrarily chosen.

\begin{figure*}
\center
\mbox{
\subfigure{\label{fig:diff_q1_q2}\epsfig{figure=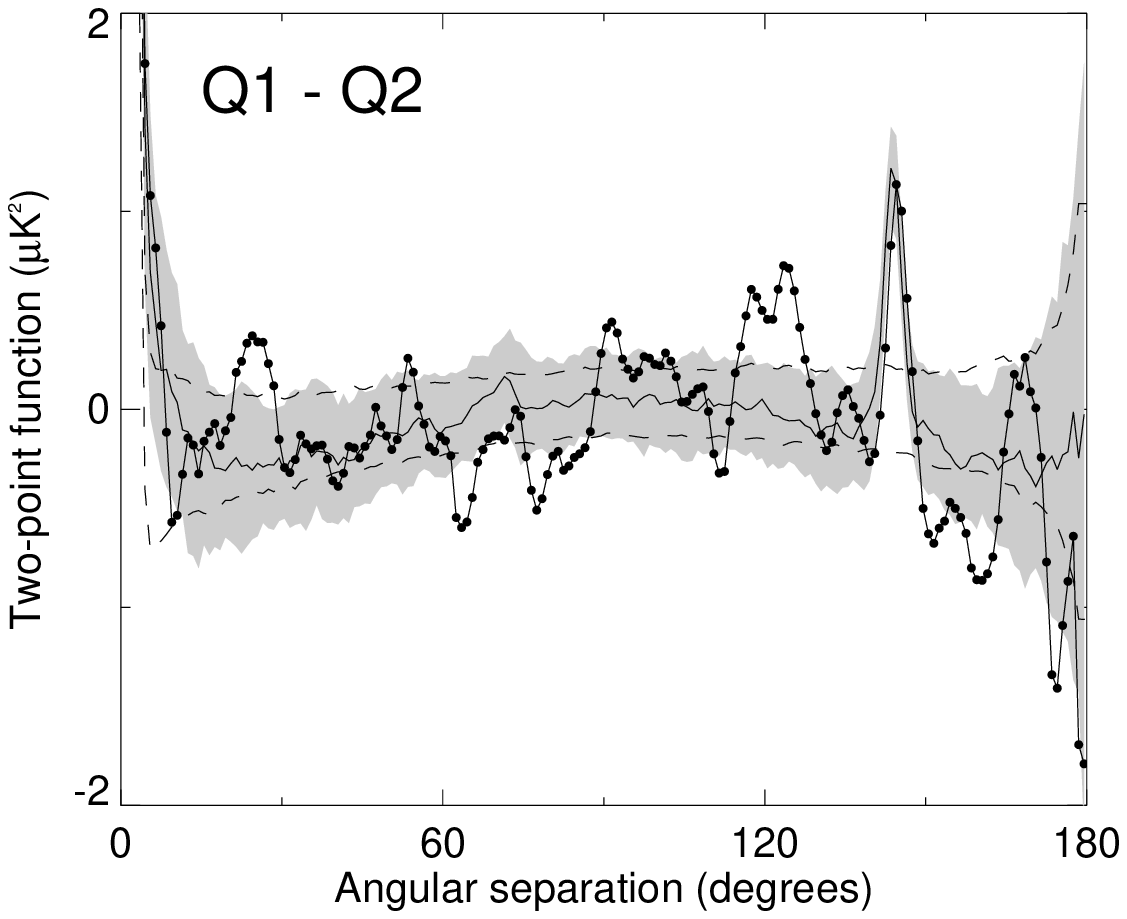,width=0.48\textwidth,clip=}}
\subfigure{\label{fig:diff_v1_v2}\epsfig{figure=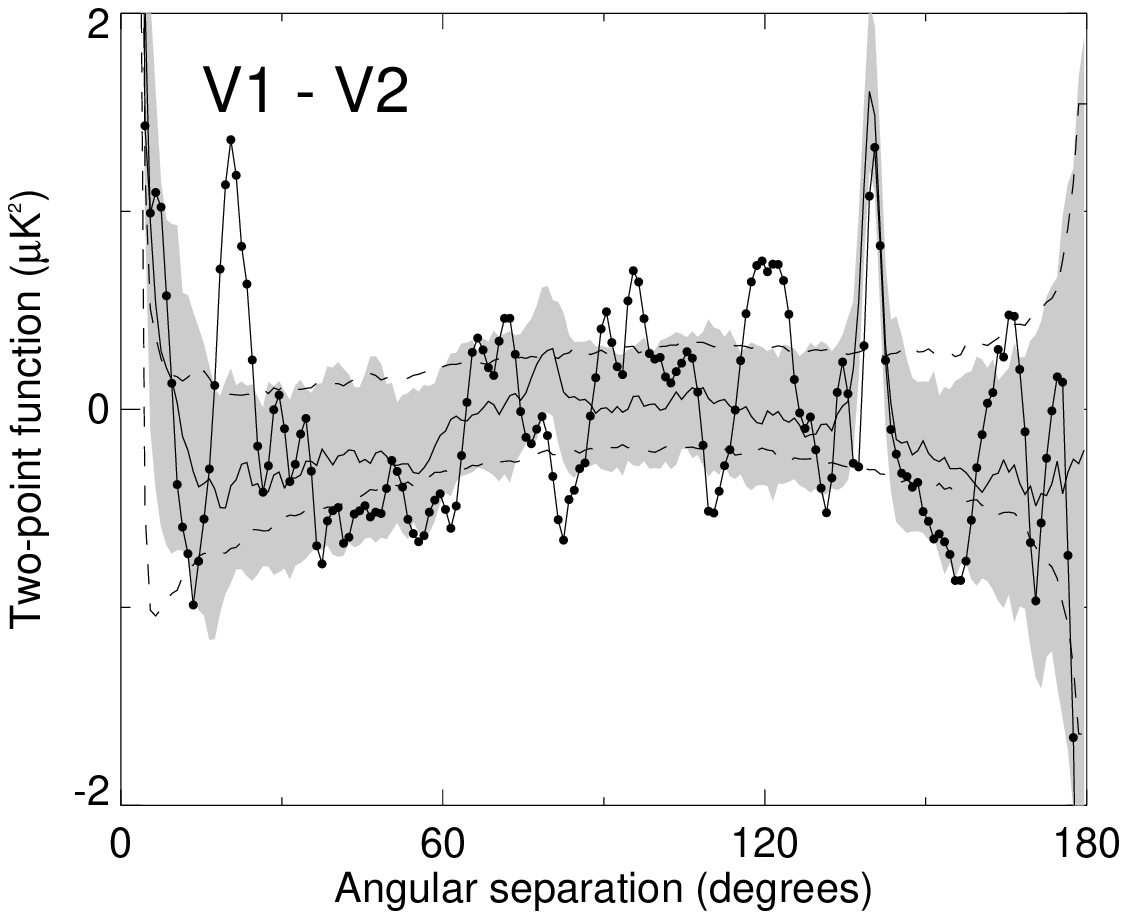,width=0.48\textwidth,clip=}}
}

\mbox{
\subfigure{\label{fig:diff_w1_w2}\epsfig{figure=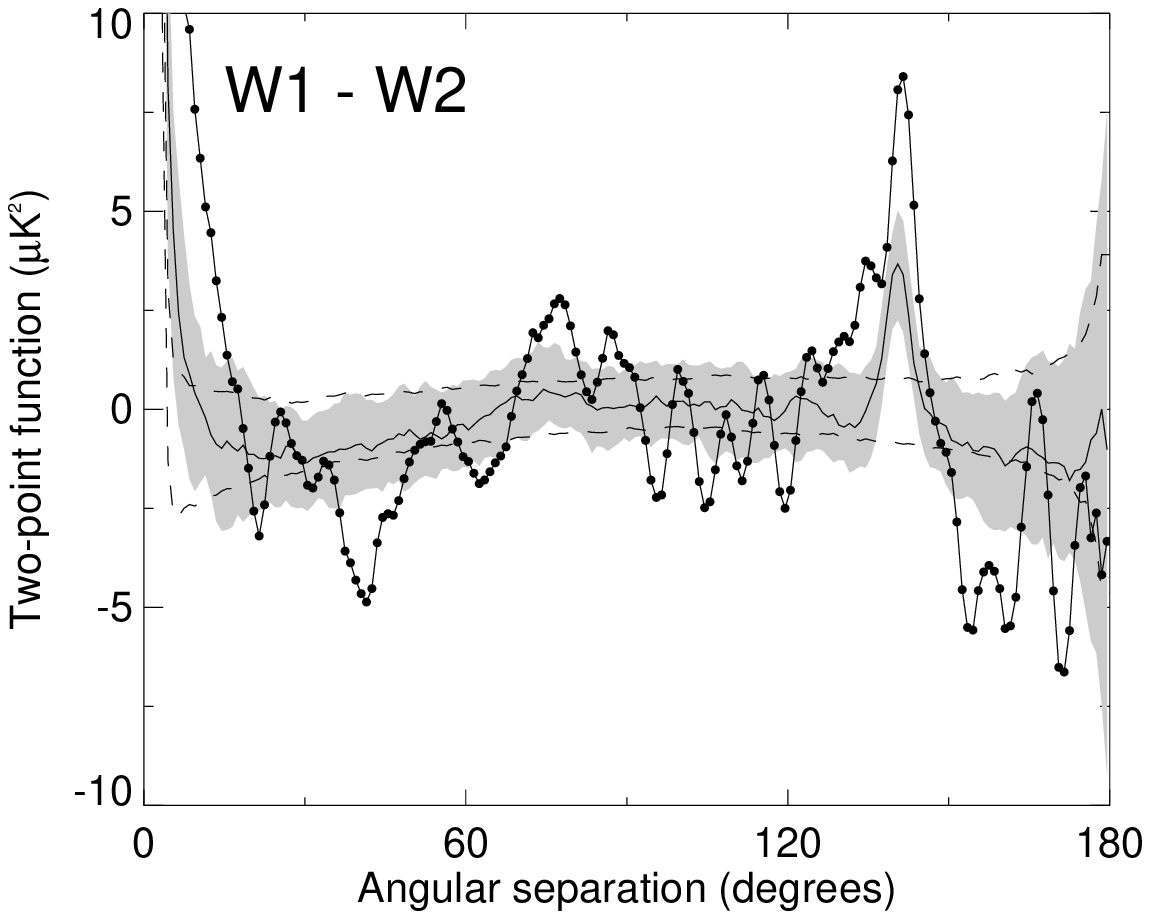,width=0.31\textwidth,clip=}}
\subfigure{\label{fig:diff_w1_w3}\epsfig{figure=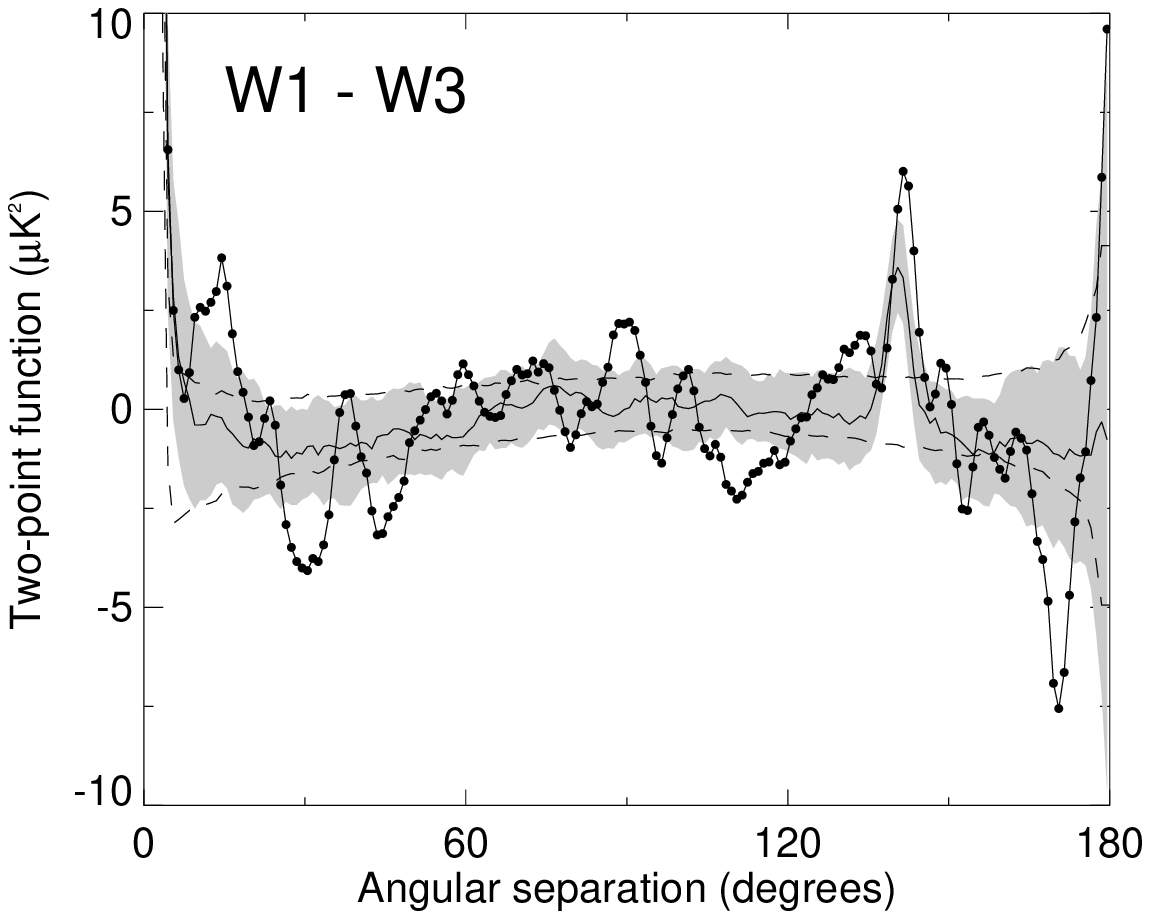,width=0.31\textwidth,clip=}}
\subfigure{\label{fig:diff_w1_w4}\epsfig{figure=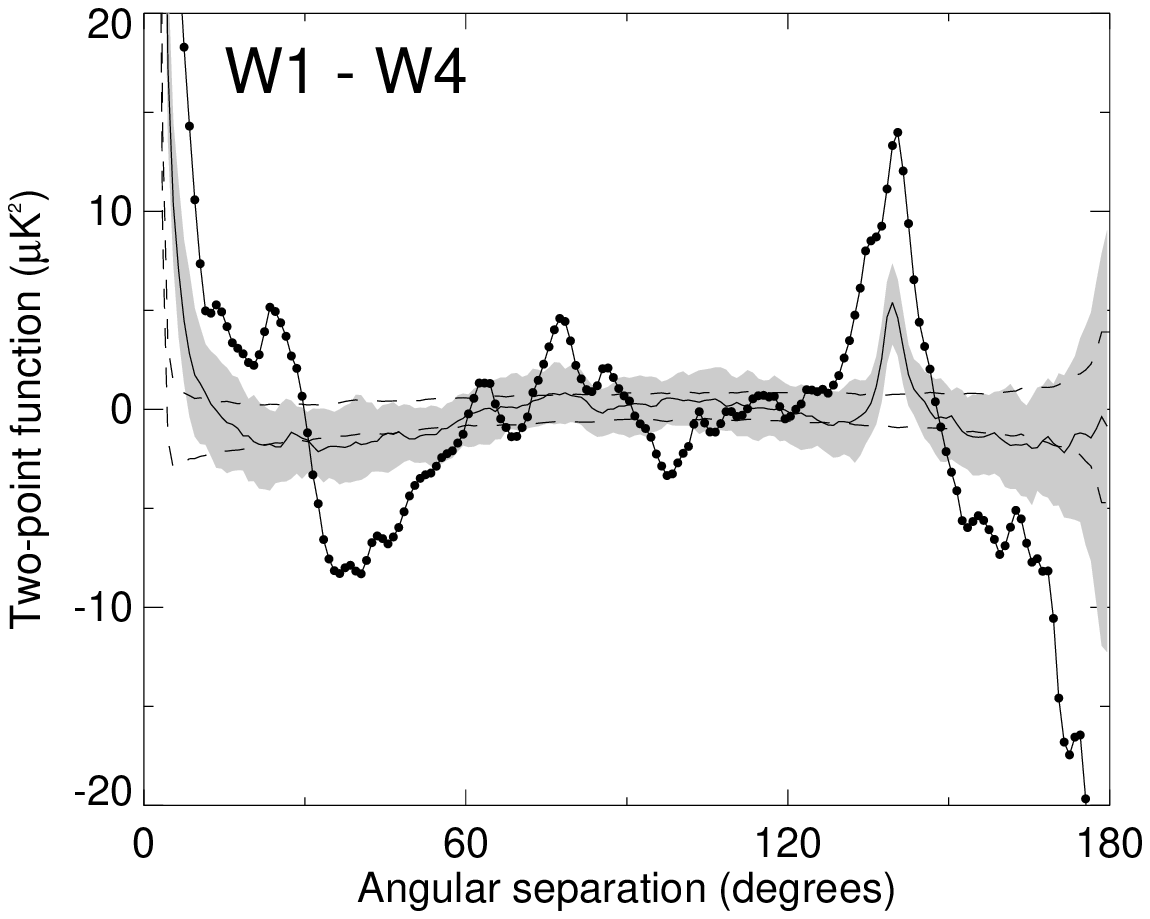,width=0.31\textwidth,clip=}}
}

\mbox{
\subfigure{\label{fig:diff_w2_w3}\epsfig{figure=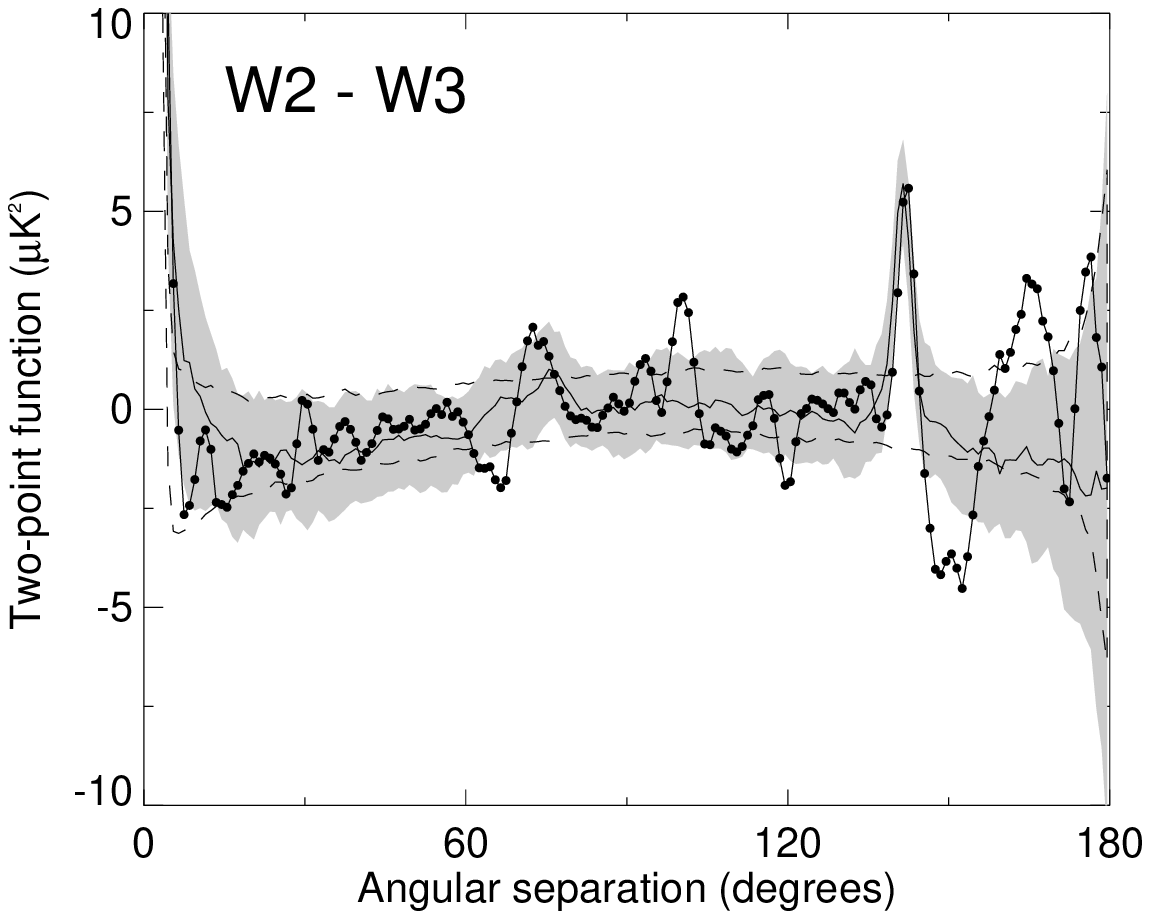,width=0.31\textwidth,clip=}}
\subfigure{\label{fig:diff_w2_w4}\epsfig{figure=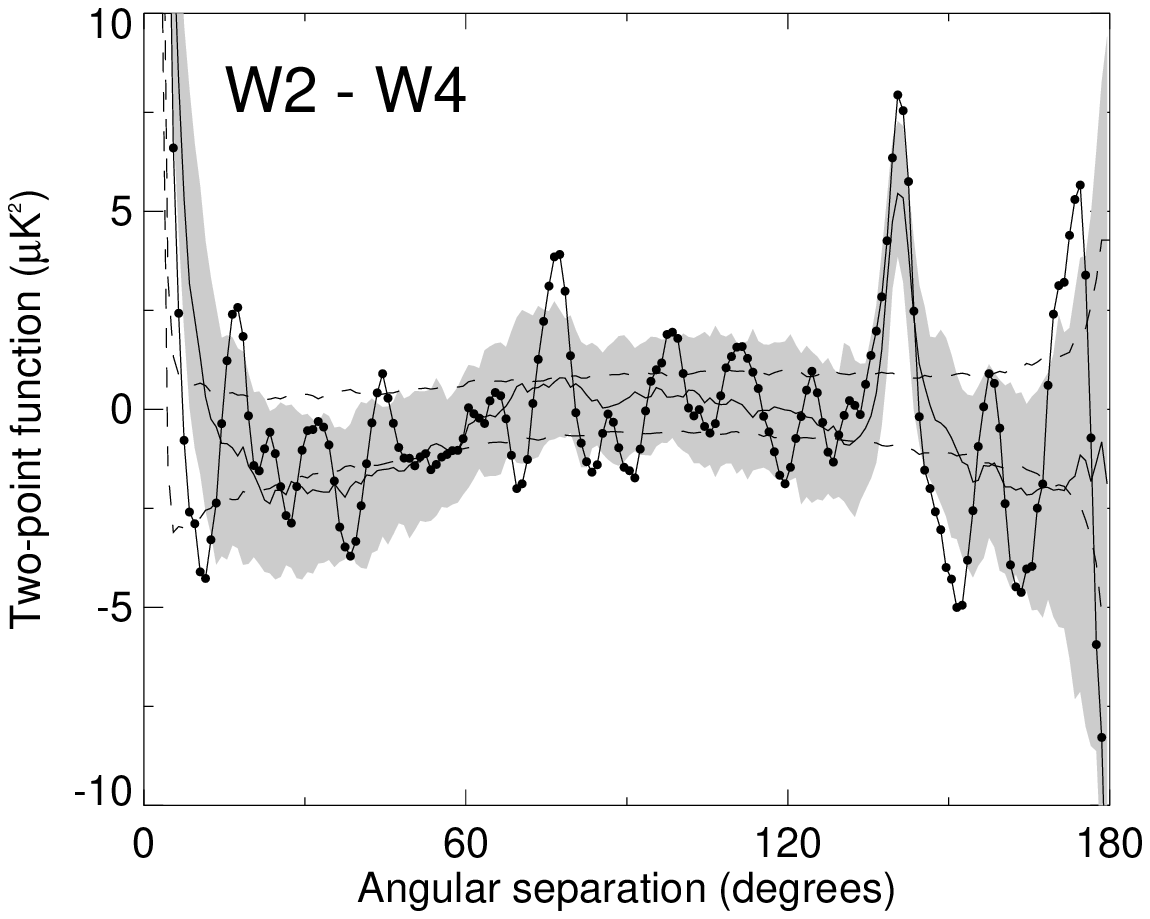,width=0.31\textwidth,clip=}}
\subfigure{\label{fig:diff_w3_w4}\epsfig{figure=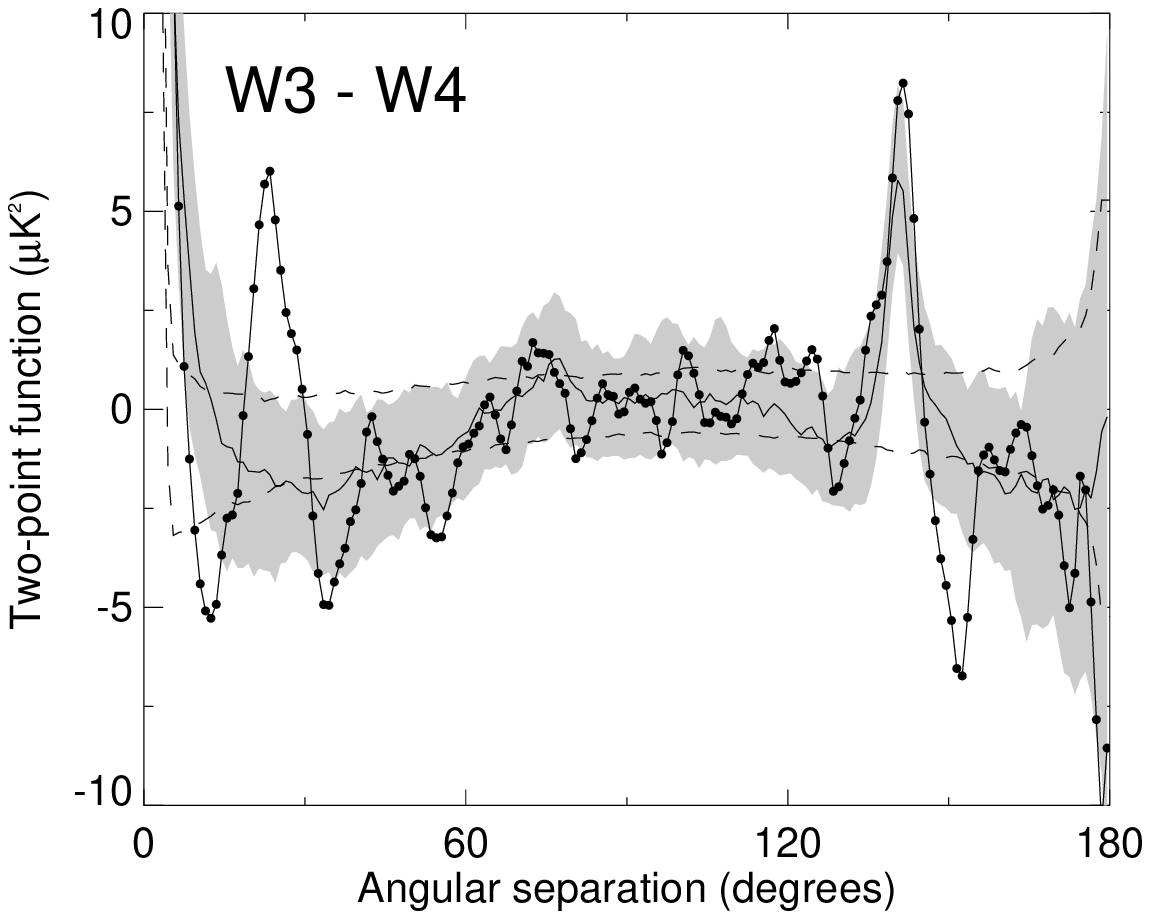,width=0.31\textwidth,clip=}}
}
\caption{Results from the difference map analysis. The solid dots show
the results computed from the observed maps, while the solid line and
the gray
bands show the median and the $1\,\sigma$ confidence region, respectively,
computed from the 110 simulations produced by the \emph{WMAP} team,
including all known systematic effects. The dashed lines indicate the
$1\,\sigma$ confidence region assuming uncorrelated noise, modulated by
$N_{\textrm{obs}}(p)$ only.}
\label{fig:noise_results}
\end{figure*}

In general, the $S$-statistic has similar properties to a $\chi^2$
statistic that only includes diagonal terms, but it has a distinct
advantage over the latter in the case of the two-point function: while
both power deficits and excesses lead to a large $\chi^2$ (rendering
this statistic useless for probing asymmetry), the opposite is true
for the $S$-statistic. Power deficit yields a low $S$-value, while
power excess yields a high $S$-value. In other words, this statistic
may serve the same purpose for the two-point function as the $\chi^2$
statistic does for the higher-order functions.

The results from this analysis are shown in the second section of
Table \ref{tab:chisq_large}. Overall, they are consistent with the
$\chi^2$-based three- and four-point function results, with the single
exception of the galactic $|b| > 30^{\circ}$ measurements, which do
not show any signs of asymmetry. However, in this case the two-point
function is quite poorly constrained at the largest angles because of
the limited sky coverage, and sample variance dominates the statistic.

In Figure \ref{fig:chi_squares} the histograms of the $\chi^2$ values
for the co-added simulated ensemble are plotted together with the
observed \emph{WMAP} values (both for the foreground-corrected and the
raw maps). In this Figure it is well worth noting the effect of
foregrounds, namely that the $\chi^2$ \emph{increases} if foregrounds
are present. This is both an intuitive and an important result: it is
intuitive because the $\chi^2$ statistic basically measures the amount
of deviations from the average function, and for a function with
vanishing mean such as the three-point function, it therefore
quantifies the overall level of fluctuations. By adding a
statistically independent component to the maps (residual foregrounds
in our setting), more fluctuations are introduced into the three-point
function. This observation is therefore also important, since it
implies that residual foregrounds are unlikely to explain the northern
hemisphere anomaly -- sub-optimal foreground templates would
\emph{introduce} large-scale fluctuations, rather than suppress
them. This also suggests that one could use the $\chi^2$ statistic as
defined above to fit for the template amplitudes, a possibility that
will be explored further in a future publication.

All in all, the results presented in this section seem to disfavor a
foreground-based explanation for the large-scale power asymmetry. The
variation from band to band is very small indeed, and similar signs of
asymmetry can be seen in any one of the frequencies. Further, there is
no clear dependence on the particular sky cut.

\subsection{Analysis of difference maps}

Next, we study the noise properties of the \emph{WMAP}
data. Specifically, the two-point correlation functions are computed from
all possible difference maps within each frequency channel, and
we compare these to the functions computed from corresponding correlated
noise maps\footnote{Available at http://lambda.gsfc.nasa.gov.}
produced by the \emph{WMAP} team. All data have been processed in the
same way as for the large-scale analysis; the maps are downgraded to a
common resolution of $140'$, then the difference maps are formed, and
finally, the best-fit monopole, dipole and quadrupole moments are
removed from the extended Kp0 region.

In Figure \ref{fig:noise_results} the results from this
analysis are shown, comparing the \emph{WMAP} data (solid dots) to the
simulations, that includes all known systematics. The gray bands
indicate the $1\,\sigma$ confidence bands computed from the 110
available simulations, and the dashed lines show the $1\,\sigma$ regions
assuming uncorrelated (but inhomogeneous) white noise, computed from
1000 simulations. 

In particular two features stand out in these plots: First, there is a
strong peak at $\theta_{\textrm{h}} = 141^{\circ}$, the
effective horn separation angle of the \emph{WMAP} satellite. Second,
there is a clear rise toward high values at small angles, which is a
real-space manifestation of correlated noise. Of course, neither of
these effects are unexpected, since they are also present in the
simulations, and they are both discussed at some length by
\citet{hinshaw:2003b}.

However, there are a few surprises to be found in the $W$-band
plots. Specifically, a very strong signal may be seen in the $W$1--$W$4
map. To the extent that the confidence regions can be approximated by
Gaussians, we see that the peak at $\theta_{\textrm{h}}$ extends to
more than $4\,\sigma$ compared to the simulations, and the overall
fluctuation levels are clearly stronger than what is seen in the
simulations. A similar pattern is also seen in the $W$1--$W$2 map, but
with a slightly smaller amplitude.

Comparing the confidence regions estimated from correlated and white
noise simulations, we see that the main difference is more large-scale
curvature in the correlated noise bands. This is consistent with the
power spectrum view, where correlated noise is found to have the
strongest impact at low $\ell$'s. This again translates into a
two-point function with a shape resembling that of the
signal-dominated functions shown in Figure
\ref{fig:corrfuncs_large}. This effect is particularly evident in the
maps which involve the W4 differencing assembly, which is known to
have a significantly higher knee frequency than the other DA's
\citep{jarosik:2003}.

We now quantify the agreement between the observations and the model
by means of a $\chi^2$ statistic, but we do not attempt to include the
correlation structure in this case because of the limited number of
simulations. Rather, we define a simplified statistic on the following
form,
\begin{equation}
\chi^{2}_{\textrm{diag}} = \sum_{i=1}^{N_{\textrm{bin}}}
\frac{\left(C_2(i)-\bigl<C_2(i)\bigr>\right)^2}{\sigma^{2}(i)}.
\end{equation}
Here $N_{\textrm{bin}}$ is the number of bins in the correlation
function, and $\bigl<C_2(i)\bigr>$ and $\sigma^{2}(i)$ are the average
and variance, respectively, of bin number $i$ computed from the
simulations. Unsurprisingly, this statistic strongly rejects the model
for the $W$1--$W$2 and $W$1--$W$4 combinations, as none of the 110 simulations
have a higher $\chi^{2}_{\textrm{diag}}$ value than the observation,
or even close to it. For the remaining six combinations, the ratios of
simulations with a lower $\chi_{\textrm{diag}}^2$ all lie comfortably
in the range between 0.28 and 0.94.

It is difficult to make firm conclusions about the origin of these
structures based on this simple analysis alone, but it is evident that
the noise simulations do not fully capture the nature of the data. 
%
On the other hand, it is also very unlikely that this effect has any
significant impact on the cosmological results from the first-year
\emph{WMAP} data release, given its relatively small amplitude. It may
be important with respect to the second-year polarization data.

\begin{figure*}

\mbox{\epsfig{file=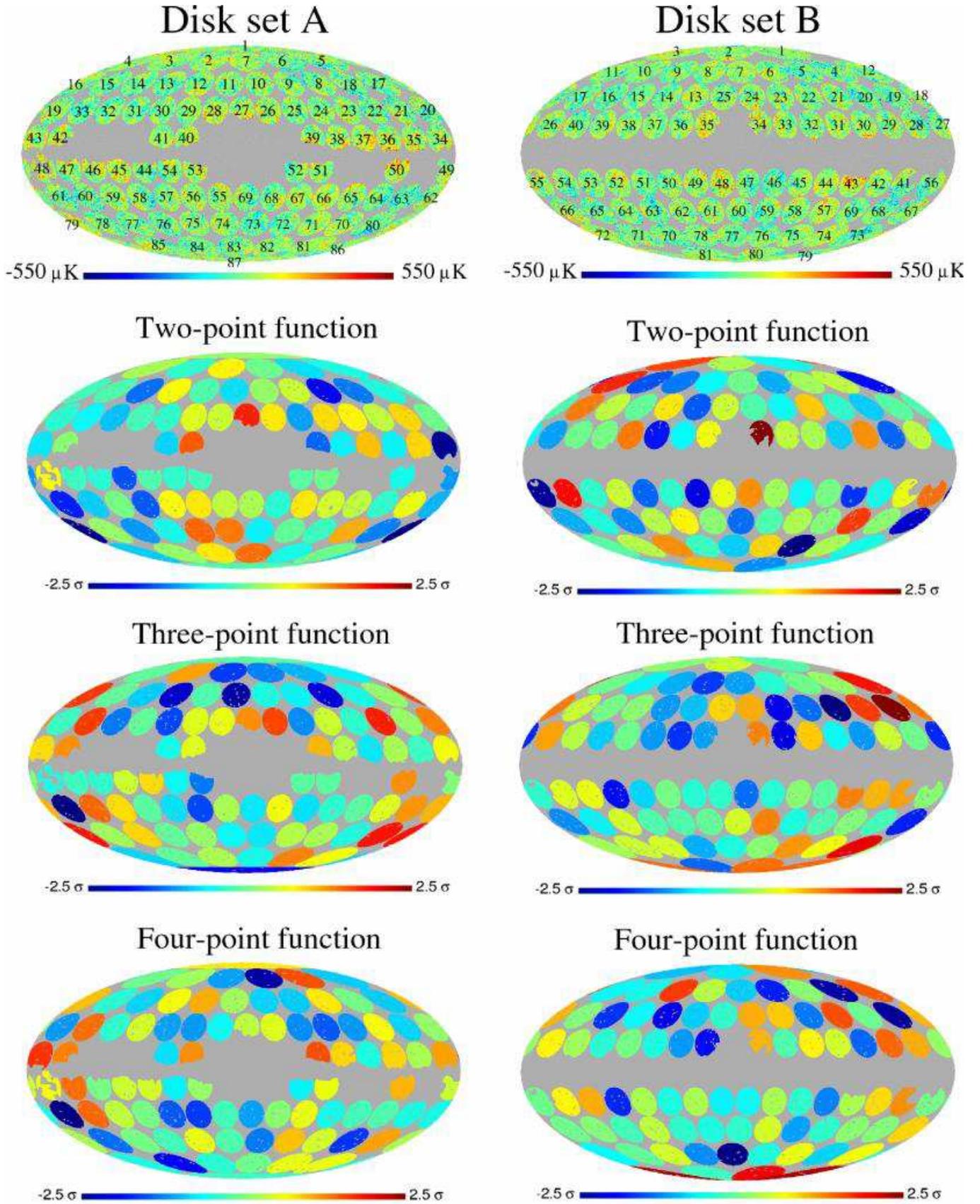,width=\linewidth,clip=}}

\caption{\tiny{Results from the intermediate scale correlation function
  analysis. The top panels show the layout of each disk set, and the
  other three rows show the $\chi^2$ results. The colors indicate the
  confidence level at which the disk is accepted, computed according
  to equation \ref{eq:gaussianize2}. Thus, dark blue indicates a very
  low $\chi^2$ value, green a value around the median, and dark red a
  very high $\chi^2$ value.}}
\label{fig:intermed_results}
\end{figure*}

A similar detection was reported by \cite{fosalba:2004}. They found
that the noise contribution in the \emph{WMAP} data may have been
underestimated by 8--15\% in the original analysis. However, it is
difficult to establish a direct connection between these results,
considering that their results are most significant at high $\ell$'s,
while our analysis is explicitly restricted to low $\ell$'s.

\section{Small- and intermediate-scale analysis}
\label{sec:small_intermed}

In order to probe smaller scales,  subsets of the
$N$-point functions are now from the full-resolution co-added map. This
analysis is facilitated by partitioning the sky into non-overlapping
disks of $10^{\circ}$ radius, each including between 15\,000 and
25\,000 pixels (the number varies because of the Kp2 mask). Two
different sets of disks (denoted A and B) are used in the following
analysis, their union covers a total of 81\% of the sky, and
about 60\% each. The two sets contain respectively 87 and 81 disks.

The reasons for dividing the sky into patches are two-fold: First,
the computational cost soon becomes difficult to handle for data sets
with more than about 150\,000 pixels. Since the algorithms scale as a
relatively high power of $N_{\textrm{pix}}$, it is much cheaper to
divide the full region into patches. Second, and equally important,
we want to be able to localize interesting effects in pixel space. In
particular, we seek to study the effects discussed in \S
\ref{sec:large_scales} further, and one convenient way of doing this
is by analyzing the sky in patches. A similar analysis was carried out
by \citet{hansen:2004b}, using a power spectrum based statistic.

The disk sets are created as follows: In each set, the disks are laid
out on rings of constant latitude, with as many disks on each ring as
there is space for without overlap (the polar rings of set B are
exceptions to this rule), and random initial longitude. Then 
the \emph{WMAP} Kp2 mask is applied and we keep only those ``disks'' (at this point
some have a rather peculiar geometry) with more than 15\,000 accepted
pixels. The reason for preferring the more liberal Kp2 mask over Kp0
is that we also want to study the effect of foregrounds in this
analysis.

The defining difference between set A and B is given by the latitudes
on which the disks are centered. In set A, the disks are laid out on
latitudes given by $\theta = k\cdot 20^{\circ}$, $k=0,\ldots,9$, while
the disks in set B are centered on $\theta = (k+1/2)\cdot
20^{\circ}$, $k=0,\ldots,8$. This difference implies that set A has
two rings of disks that touches the galactic plane, while the center
ring of disk set B is completely discarded. It is therefore reasonable
to assume that disk set A is more affected by foregrounds than disk
set B, as will be confirmed later. The two disk sets are shown in the
two top panels of Figure \ref{fig:intermed_results}, superimposed on
the co-added \emph{WMAP} map.

\subsection{Intermediate-scale analysis}

First, we consider the $N$-point correlation functions on intermediate
scales, here defined as scales smaller than 5--$10^{\circ}$. Each
function is binned with $7\farcm 2$ bin size, and the two-point function is
computed up to $10^{\circ}$, for a total of 83 bins. The three-point
function is computed over all isosceles triangles for which the
baseline is the longest edge, but no longer than $5^{\circ}$. Note
that this set includes the equilateral triangle and three points on a
line as special cases. Finally, the four-point function is computed
over the same set of configurations, but with a fourth point added by
reflecting the third point about the baseline. The total number of
independent configurations is about 460. Note that since there are
many more isosceles triangles with $5^{\circ}$ baseline than with
$1^{\circ}$, the vast majority of these configurations span scales
from $3^{\circ}$ to $5^{\circ}$. Consequently, the following $\chi^2$
analysis is dominated by intermediate scales rather than small scales,
even though a few small-scale configurations are included.

\begin{figure*}
\center
\mbox{\subfigure{\label{fig:hist_A_intermed_twopt}\epsfig{figure=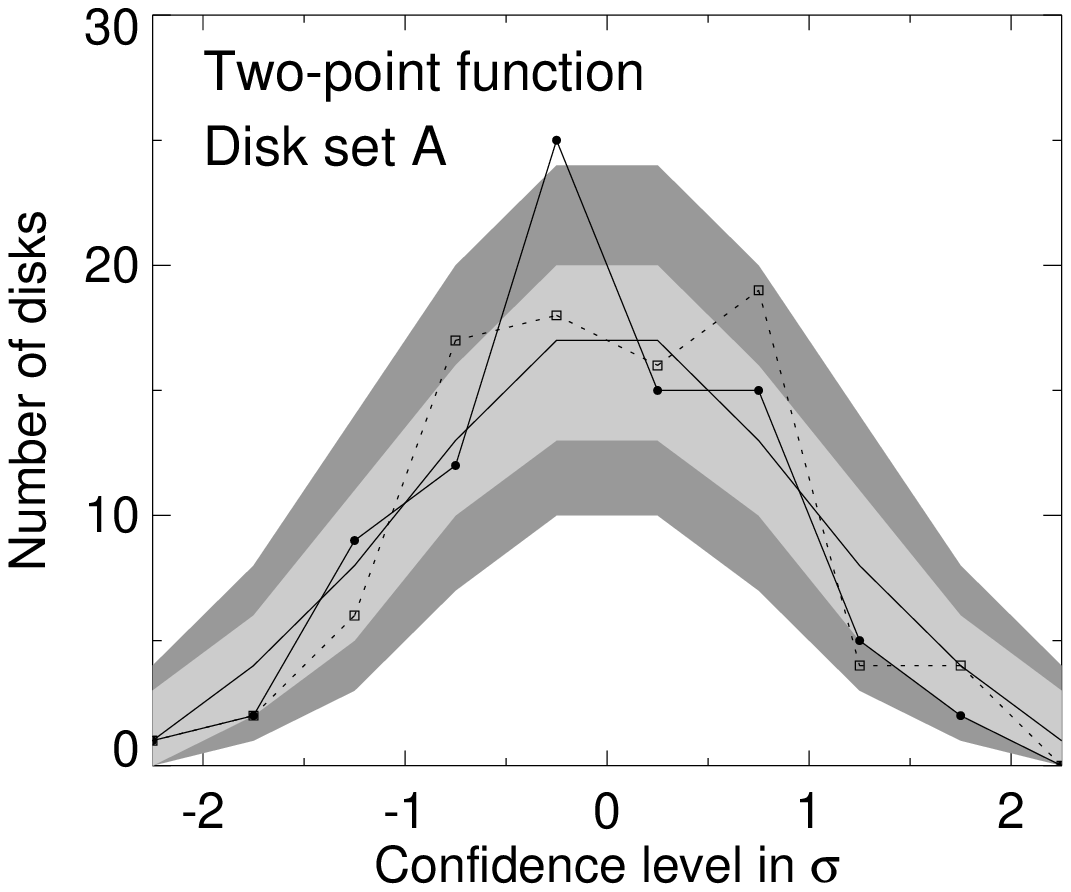,width=0.31\textwidth,clip=}}
\subfigure{\label{fig:hist_A_intermed_threept}\epsfig{figure=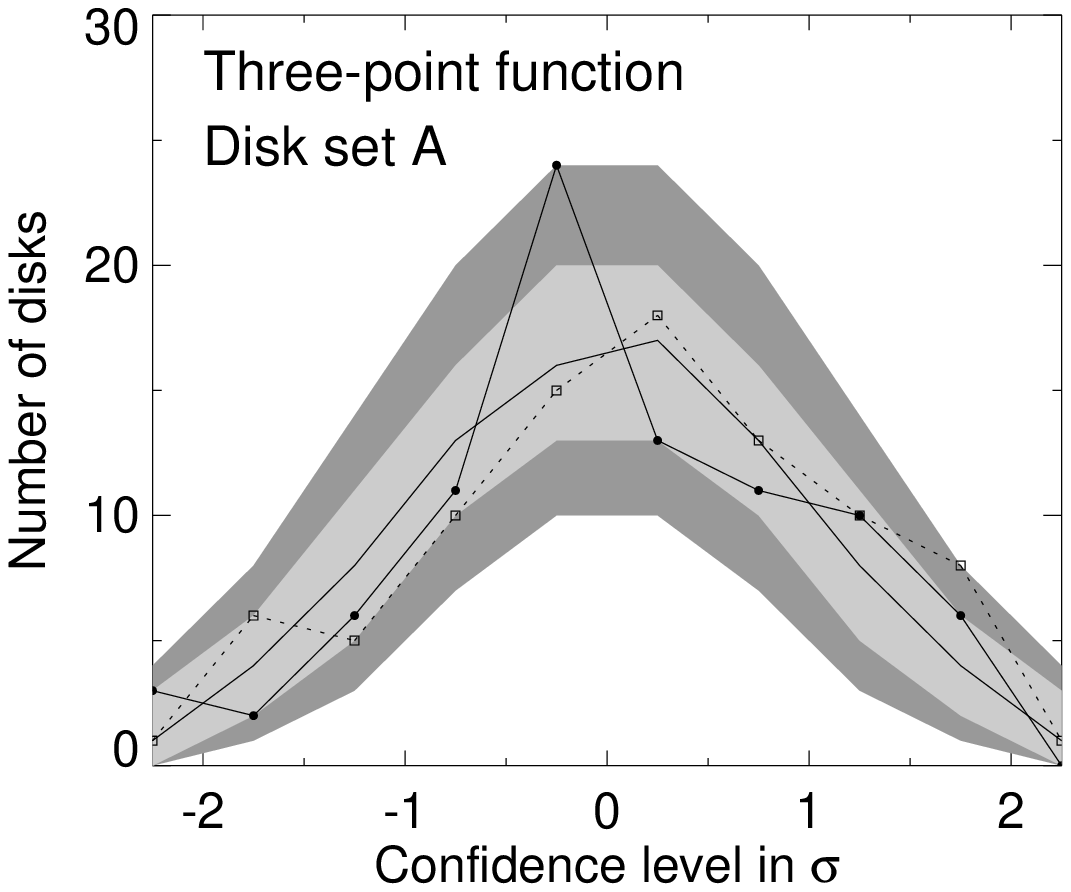,width=0.31\textwidth,clip=}}
\subfigure{\label{fig:hist_A_intermed_fourpt}\epsfig{figure=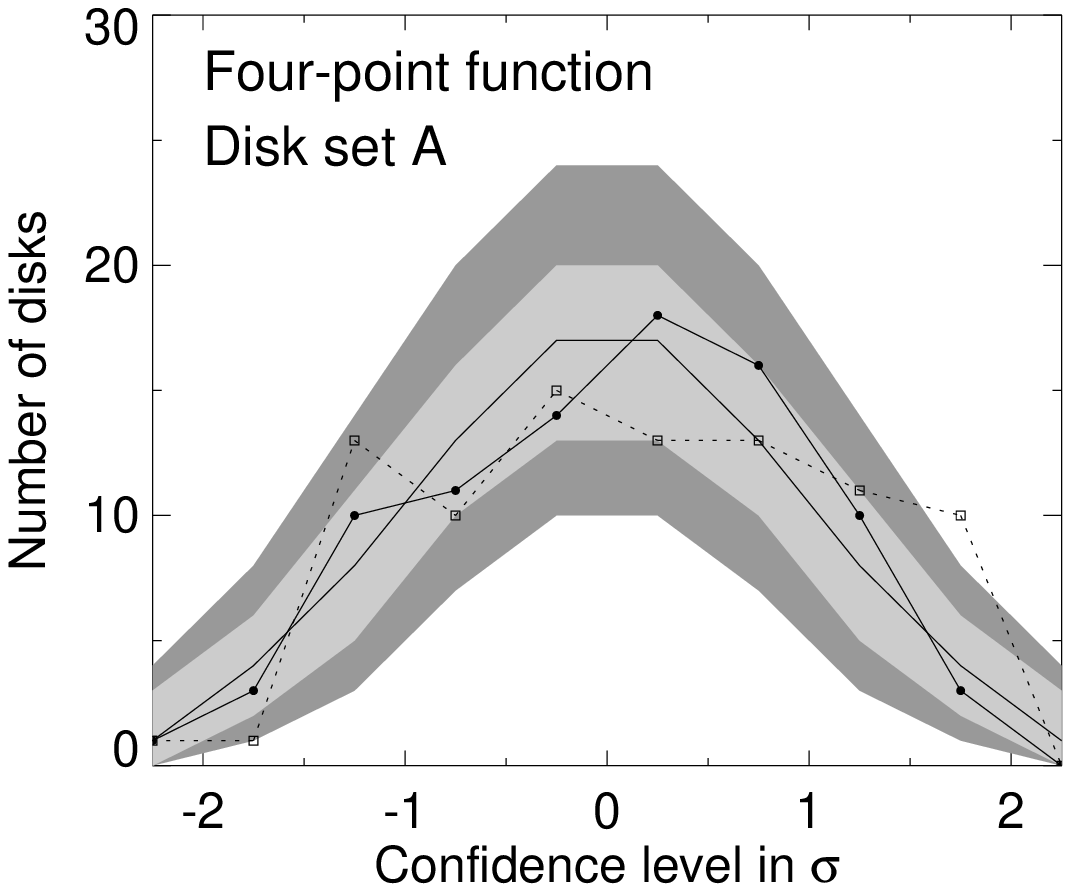,width=0.31\textwidth,clip=}}
}

\mbox{\subfigure{\label{fig:hist_B_intermed_twopt}\epsfig{figure=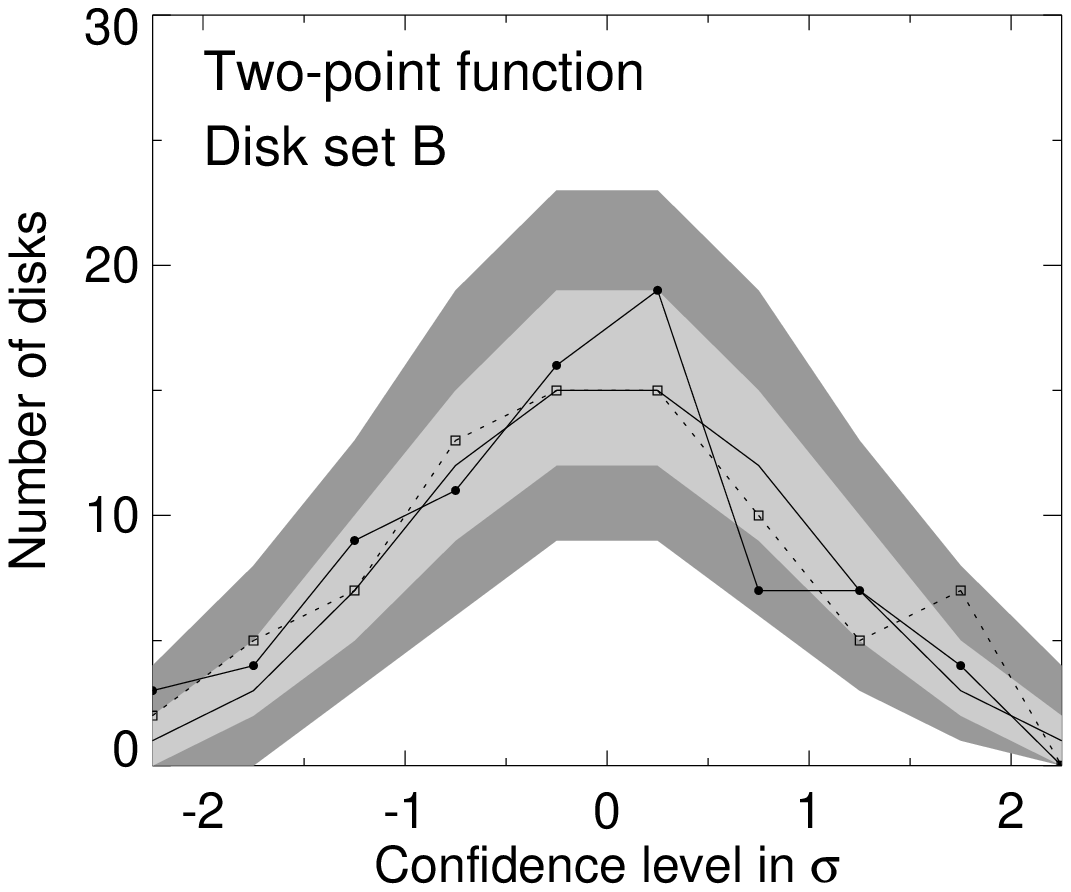,width=0.31\textwidth,clip=}}
\subfigure{\label{fig:hist_B_intermed_threept}\epsfig{figure=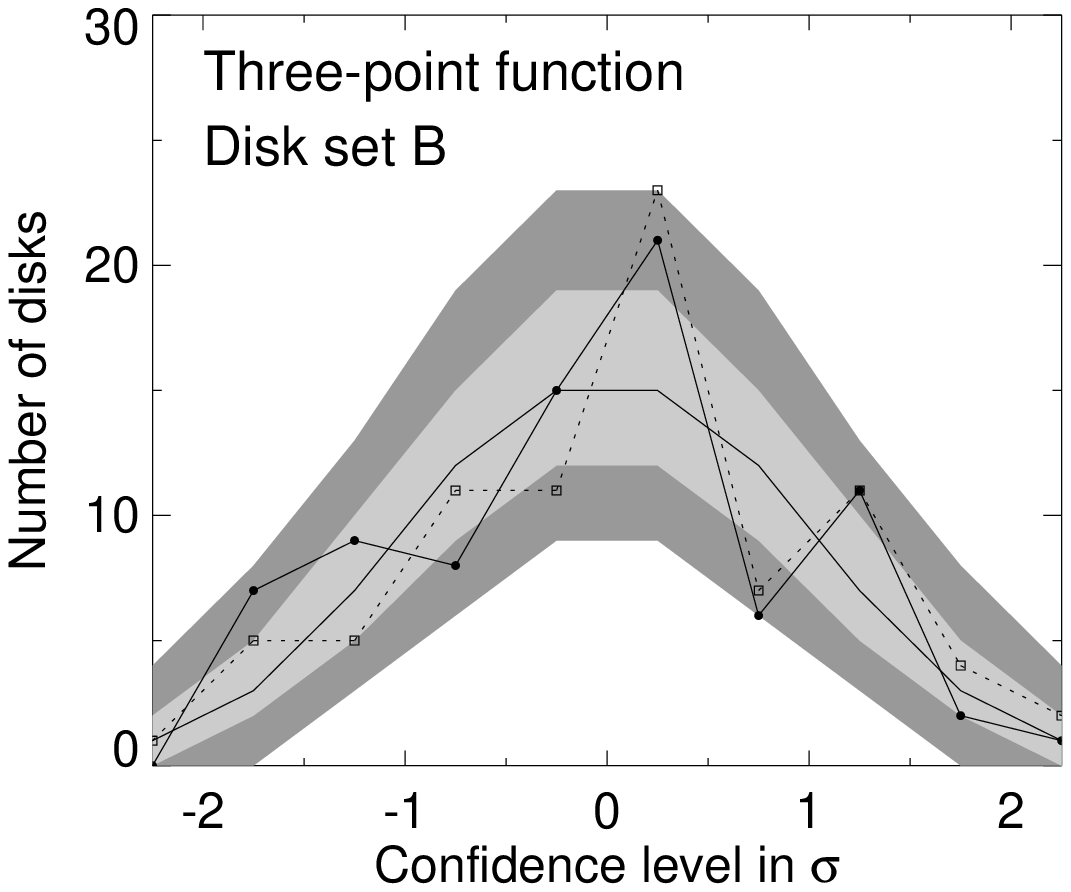,width=0.31\textwidth,clip=}}
\subfigure{\label{fig:hist_B_intermed_fourpt}\epsfig{figure=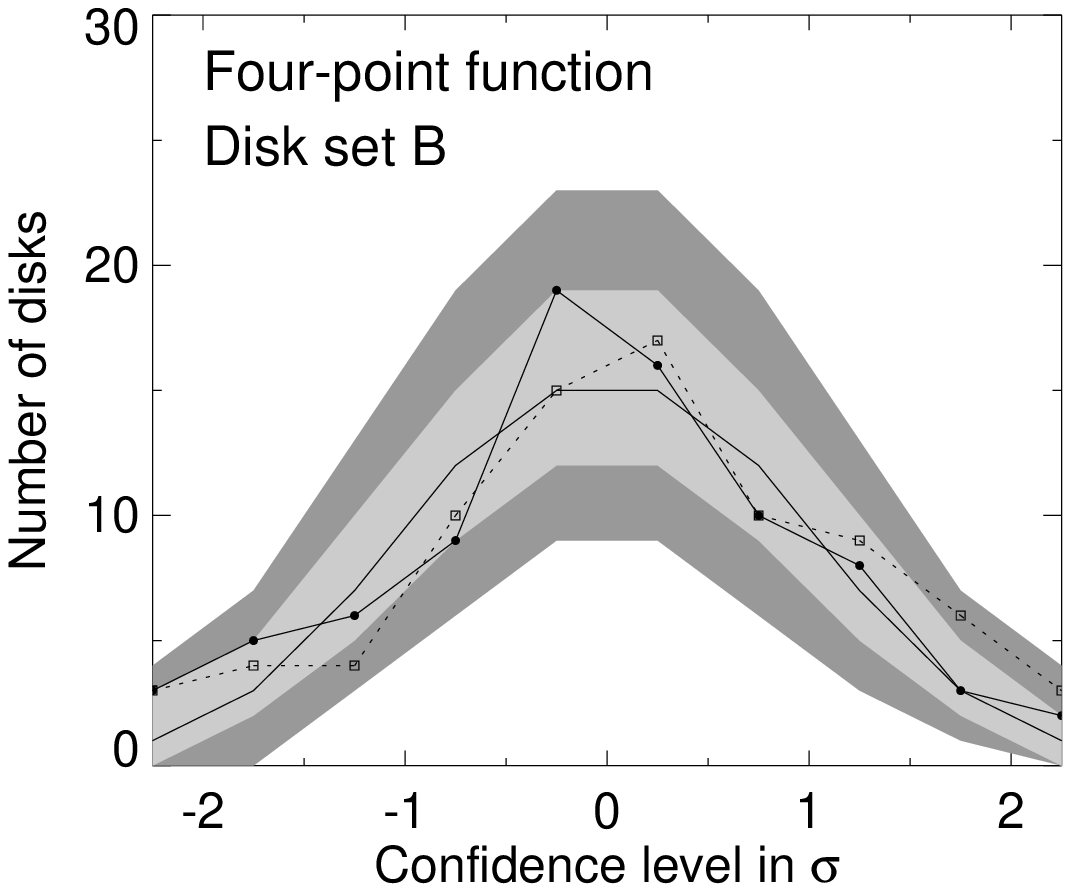,width=0.31\textwidth,clip=}}
}
\caption{Distributions (histograms) of the intermediate scale disk
  confidence levels. The top row shows the results for disk set A, the
  bottom for disk set B. The columns show, left to right, the two-,
  three-, and four-point function results. The gray bands indicate 1
  and $2\,\sigma$ confidence regions, computed from 5000
  simulations. The solid line indicates the results from the
  foreground corrected map, and the dotted line shows the results from
  the original co-added map.}
\label{fig:hist_A_intermed}
\end{figure*}

The results from this intermediate scale analysis are plotted in
Figure \ref{fig:intermed_results}, where the colors indicate the
confidence level at which each disk is accepted, as computed by
equation \ref{eq:gaussianize2}. However, extreme limits of
$-2.5\,\sigma$ and $2.5\,\sigma$ are enforced, because we only have a
limited number of simulations available. Here it is worth recalling
that we removed all power with $\ell \le 18$ from the maps, and
neighboring disks are therefore nearly uncorrelated. Distributions of
the confidence level distributions are shown in Figure
\ref{fig:hist_A_intermed}.

We see that the two-point function is to a very good approximation accepted
by the $\chi^2$ test. There are no visibly connected patches of
similar values, and the distribution of confidence values appears to
be typical compared to the simulations. The three-point
functions are more suspicious, especially when considering the pattern
seen in disk set \hbox{B}. In this case, two large, connected patches
of low $\chi^2$ values are visible on the northern hemisphere, while
the south-east quadrant appears to have quite large $\chi^2$
values. In other words, the asymmetry pattern found in the large-scale
functions is apparent even in this plot. For disk set A, these features
are less clear, but still consistent with disk set B; the northern
hemisphere on average have quite low $\chi^2$ values (or little
fluctuations), while the south-east quadrant has quite high $\chi^2$
values. This is particularly evident if one disregards all disks
touching (i.e., are partially cut by) the galactic plane, they are
more likely to be affected by residue foreground contamination.

The four-point functions are less decisive, and the general agreement
with the Gaussian model seems to be quite good. No particular features
are seen in these cases.

We now estimate the full sky correlation functions by averaging over
all the individual disk correlation functions, weighting each
sub-function by the number of pixel combinations, $N_{\textrm{c}}$,
\begin{equation}
C_N^{\textrm{full sky}}(i) \approx
\frac{\sum_{j=1}^{N_{\textrm{disks}}}
N_{\textrm{c}}^j(i)\,C_N^{j}(i)}{\sum_{j=1}^{N_{\textrm{disks}}}
N_{\textrm{c}}^j(i)}.
\label{eq:fullsky_avg}
\end{equation}
Here $C_N$ is the full-sky $N$-point correlation function, $C_N^{j}$
is the $j$'th disk correlation function, and $i$ represents the
geometric configuration under consideration.

This function is computed from each disk set individually and over the
union of the two sets. While the latter function obviously has the
advantage of larger sky coverage, it also weights configurations that
are completely contained in the intersection of two disks twice. On
the other hand, the number of such common configurations is fairly
small, at least in this intermediate-scale analysis.

\begin{figure*}
\mbox{\subfigure{\label{fig:selfunc_intermed_twopt_both}\epsfig{figure=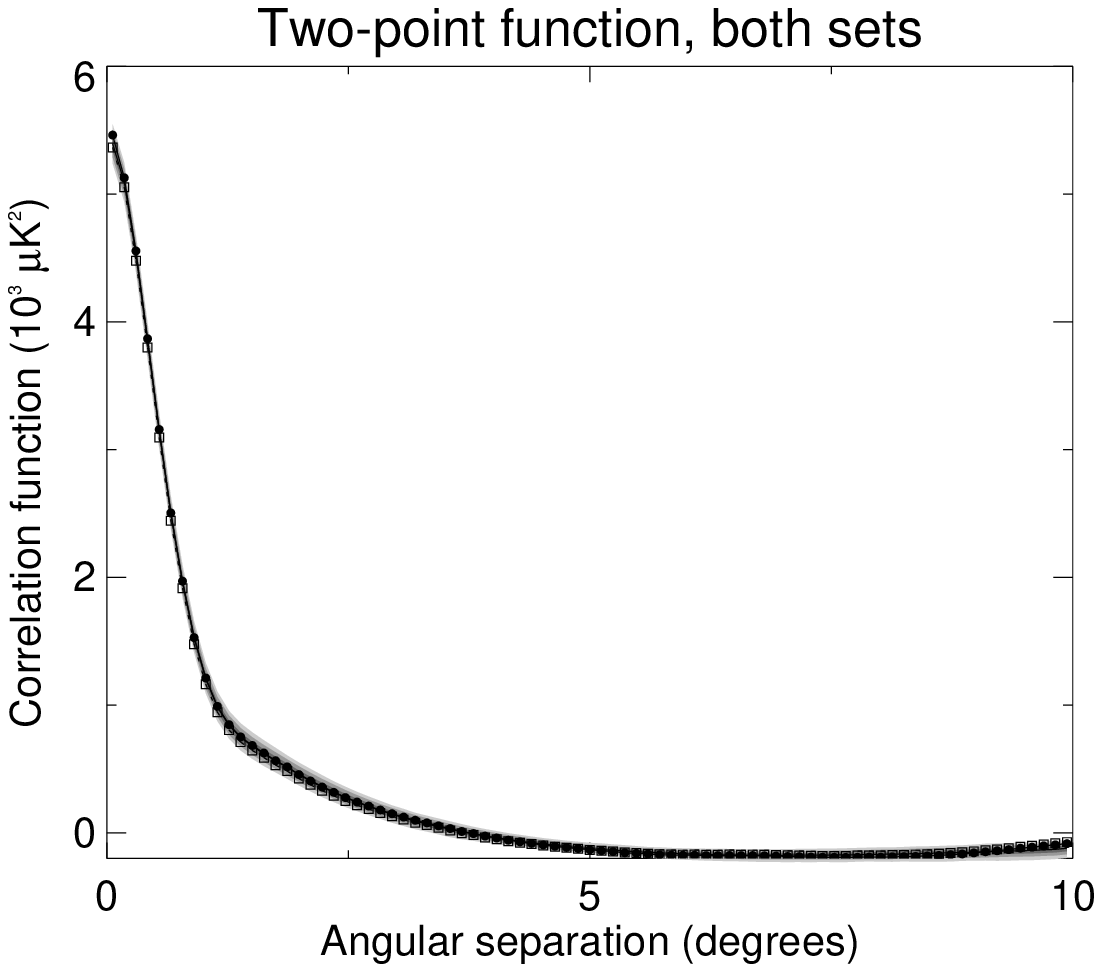,width=0.25\textwidth,clip=}}
\subfigure{\label{fig:selfunc_intermed_gauss_twopt_both}\epsfig{figure=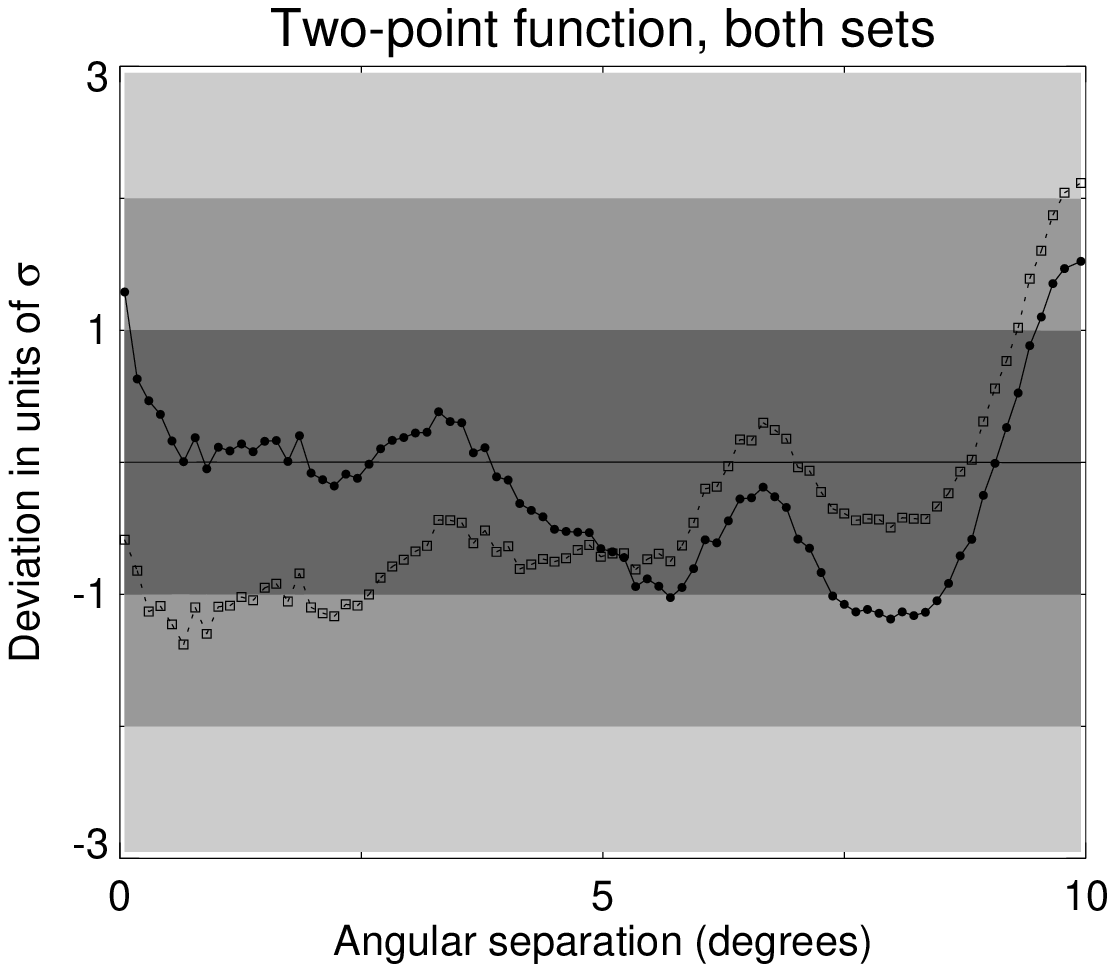,width=0.25\textwidth,clip=}}
\subfigure{\label{fig:selfunc_intermed_gauss_twopt_diskA}\epsfig{figure=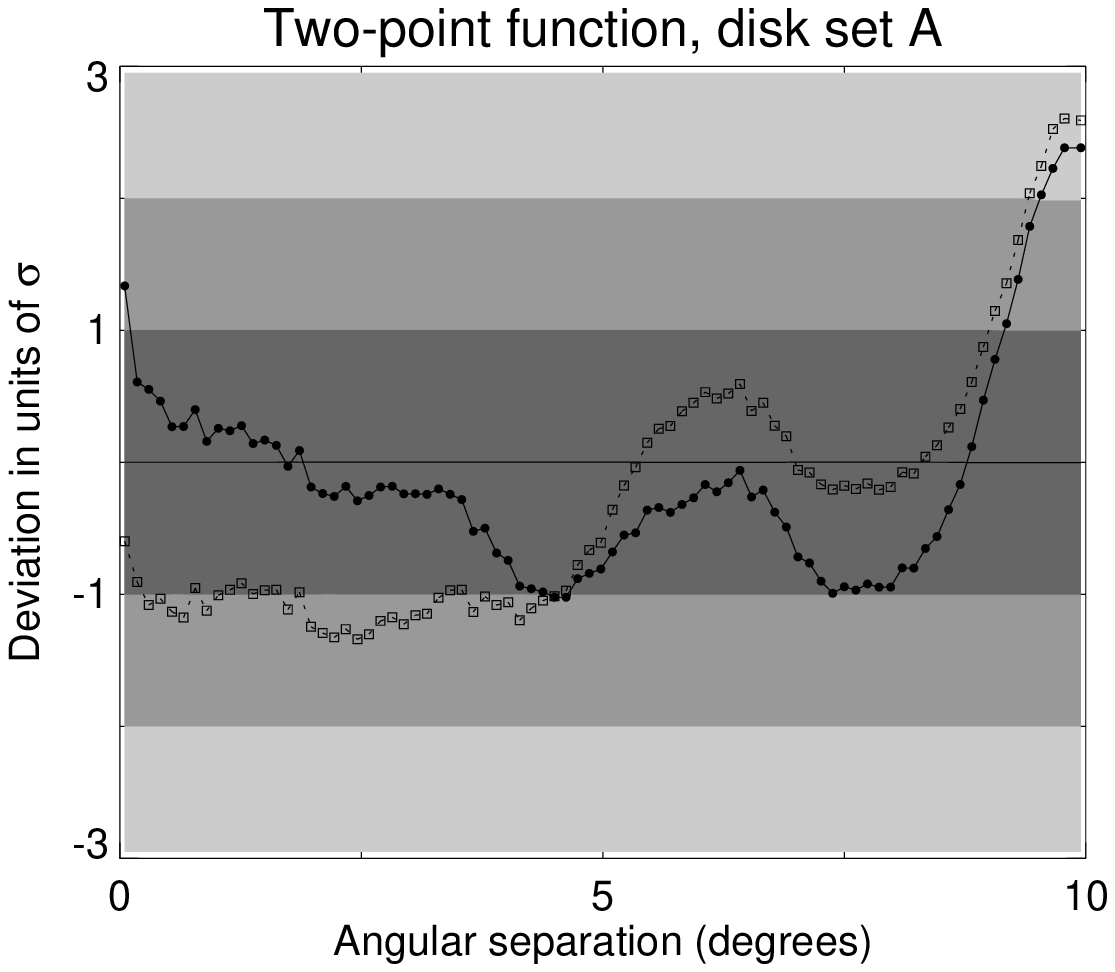,width=0.25\textwidth,clip=}}
\subfigure{\label{fig:selfunc_intermed_gauss_twopt_diskB}\epsfig{figure=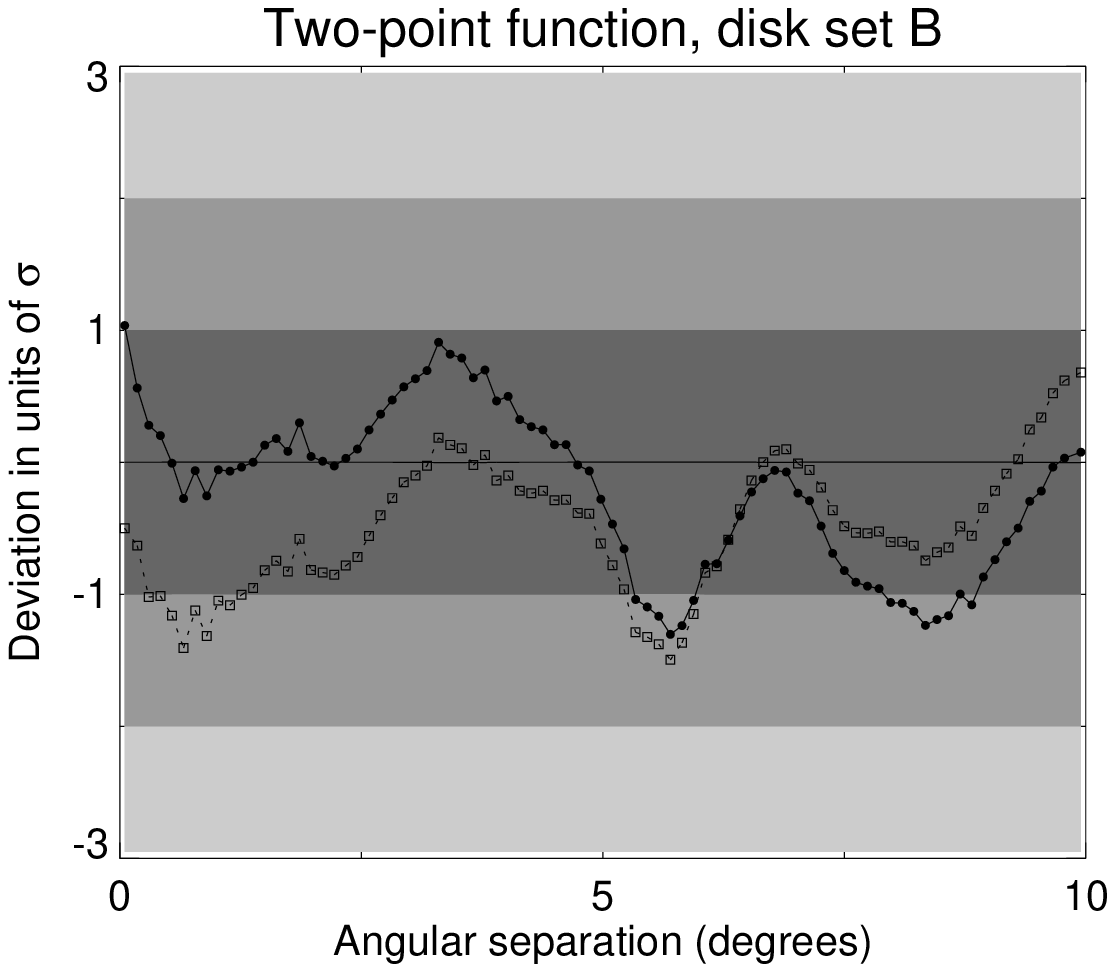,width=0.25\textwidth,clip=}}}

\mbox{\subfigure{\label{fig:selfunc_intermed_threept_both}\epsfig{figure=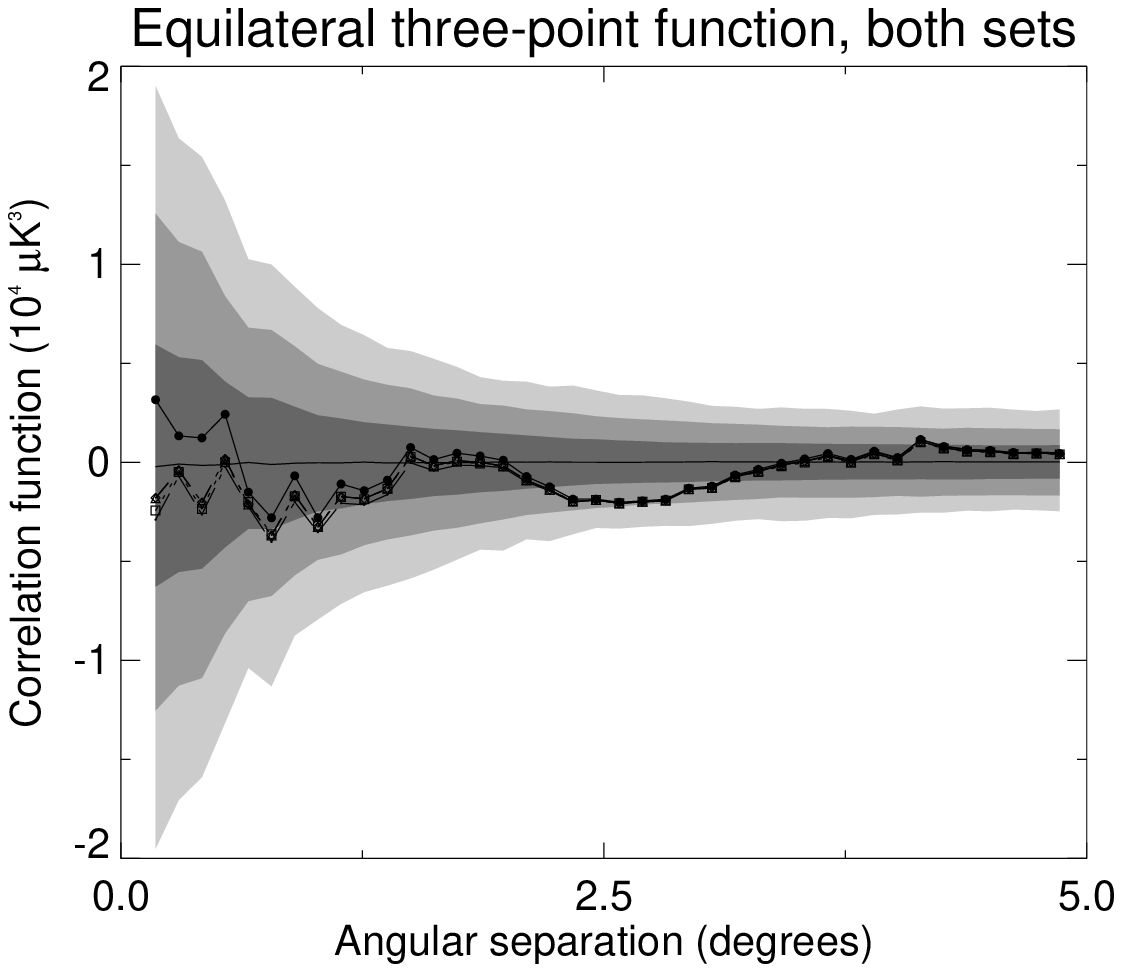,width=0.25\textwidth,clip=}}
\subfigure{\label{fig:selfunc_intermed_gauss_threept_both}\epsfig{figure=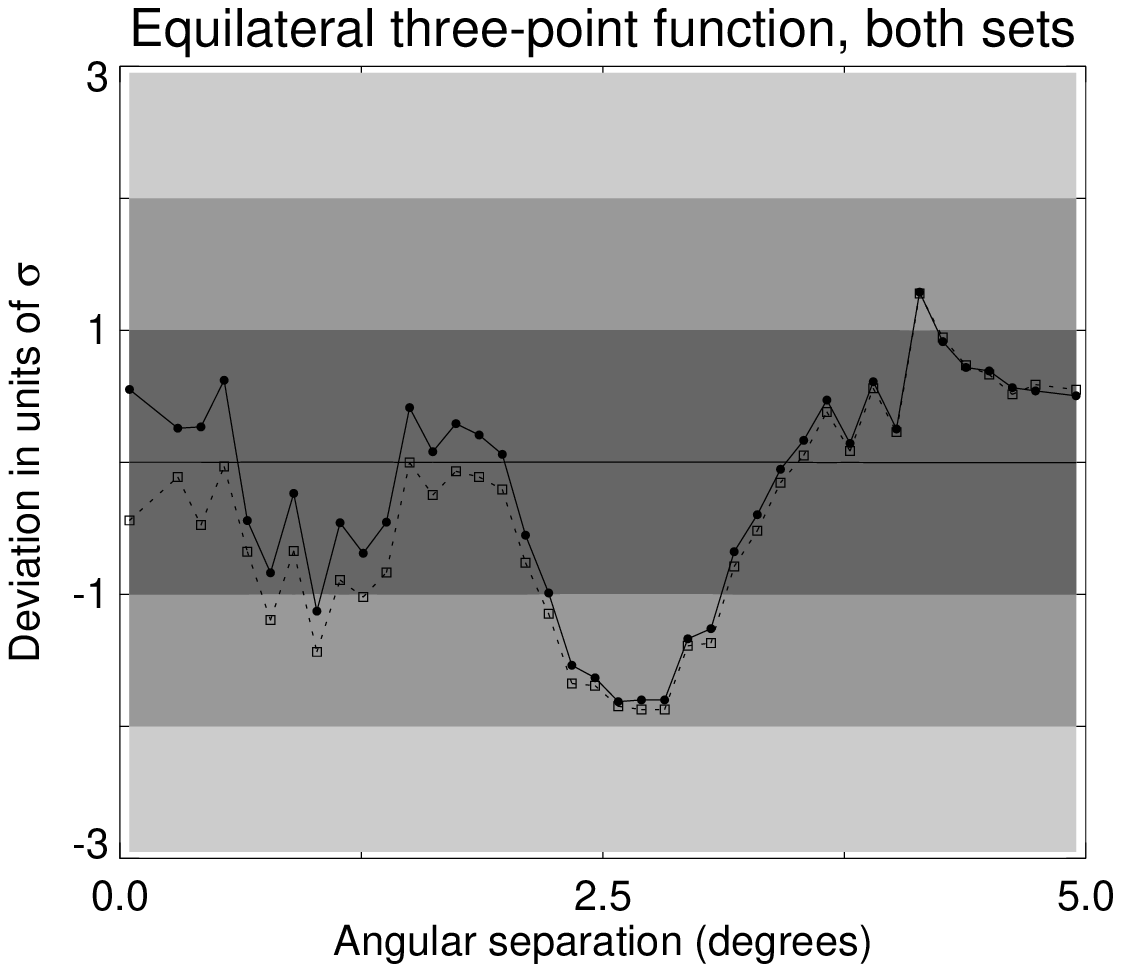,width=0.25\textwidth,clip=}}
\subfigure{\label{fig:selfunc_intermed_gauss_threept_diskA}\epsfig{figure=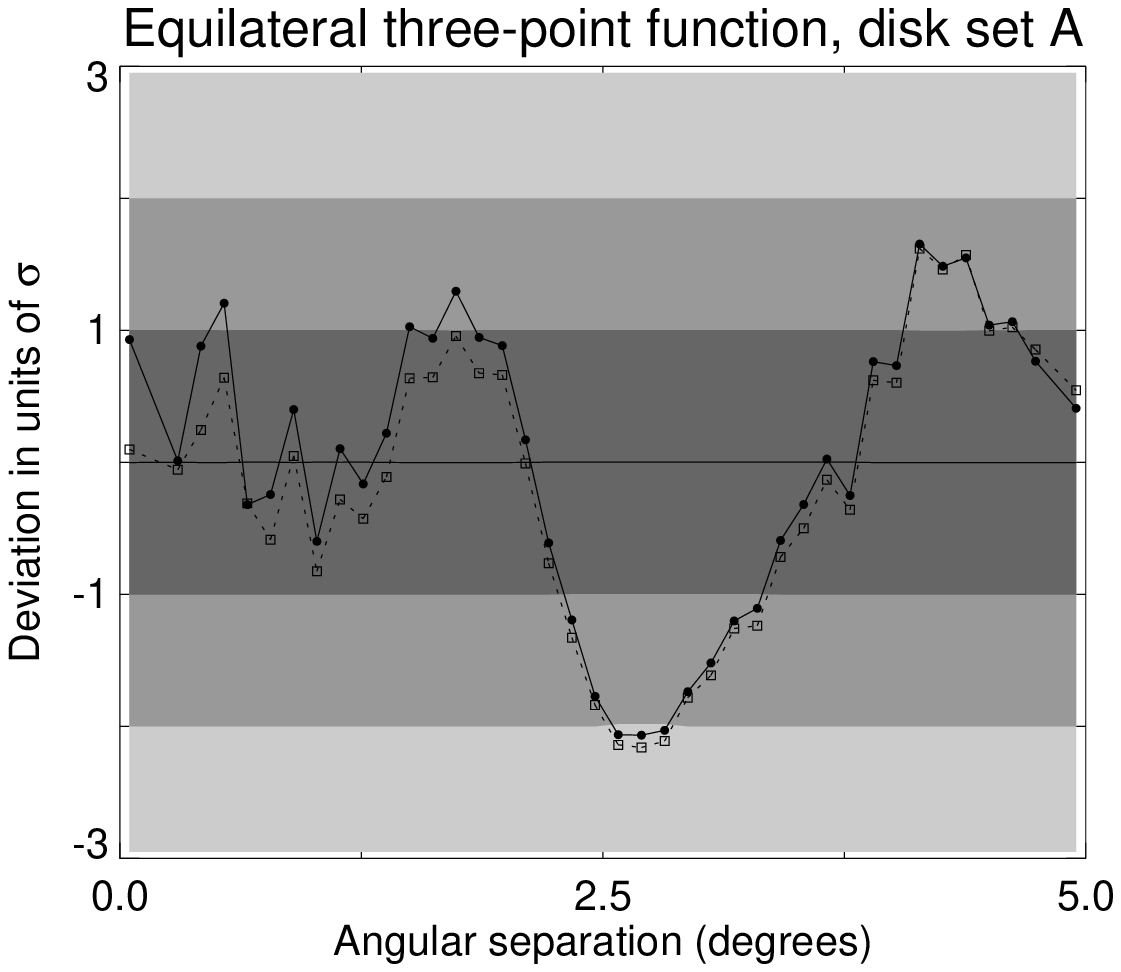,width=0.25\textwidth,clip=}}
\subfigure{\label{fig:selfunc_intermed_gauss_threept_diskB}\epsfig{figure=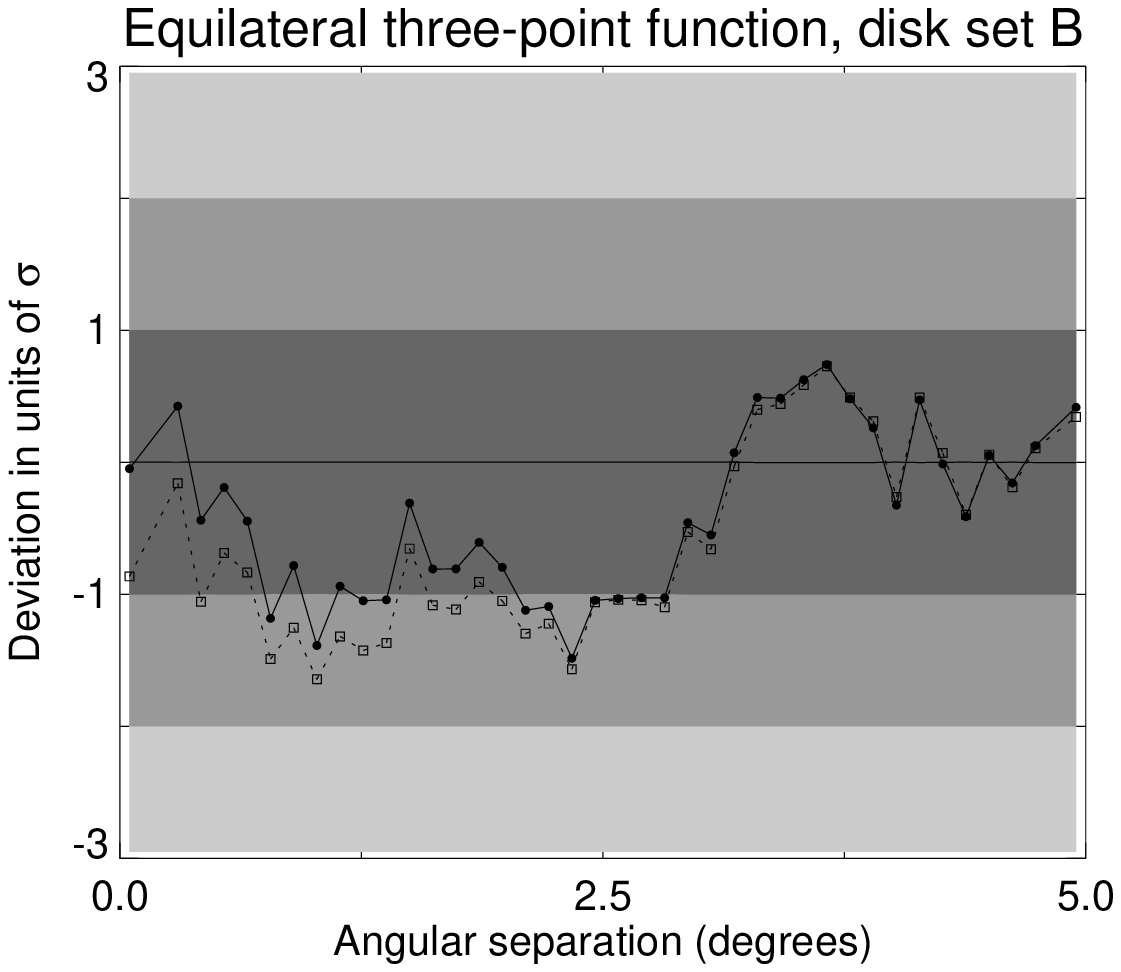,width=0.25\textwidth,clip=}}}

\mbox{\subfigure{\label{fig:selfunc_intermed_fourpt_both}\epsfig{figure=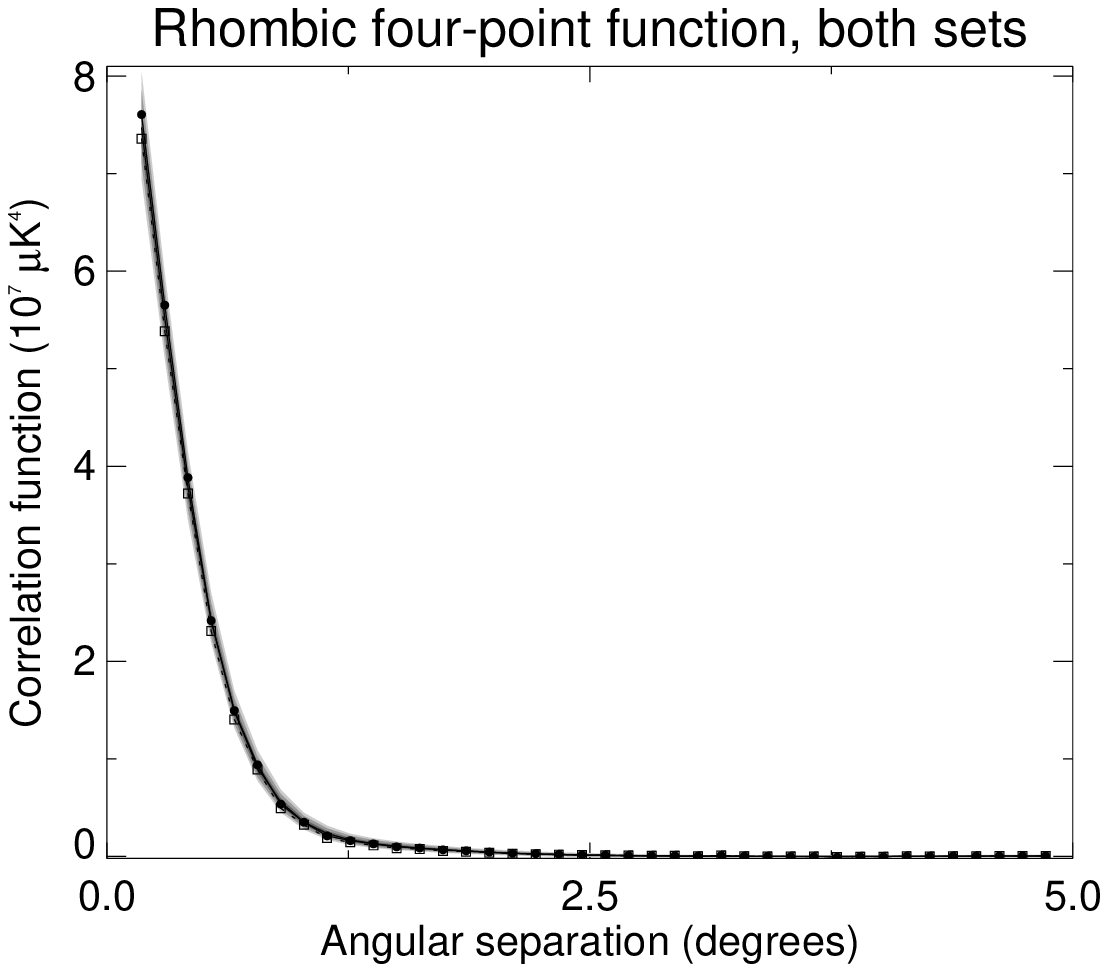,width=0.25\textwidth,clip=}}
\subfigure{\label{fig:selfunc_intermed_gauss_fourpt_both}\epsfig{figure=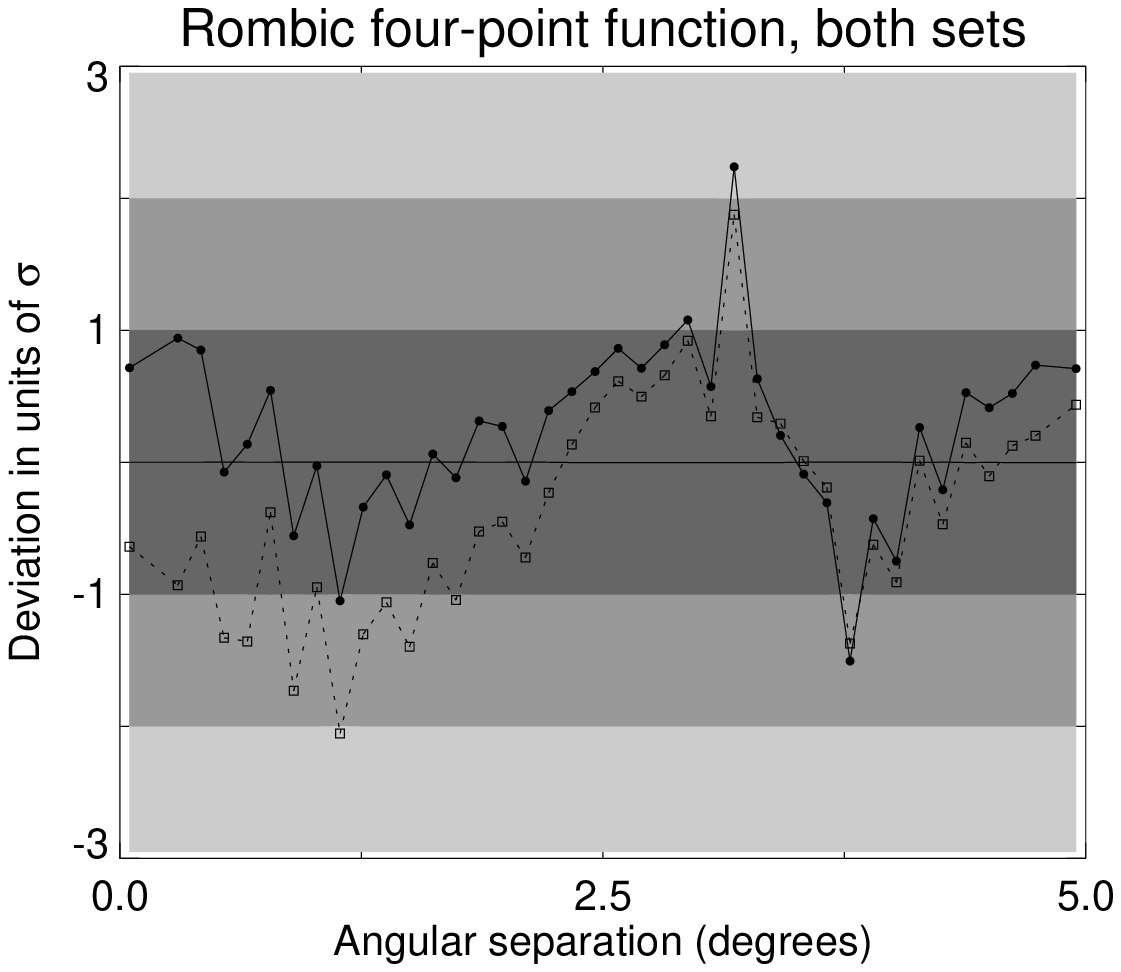,width=0.25\textwidth,clip=}}
\subfigure{\label{fig:selfunc_intermed_gauss_fourpt_diskA}\epsfig{figure=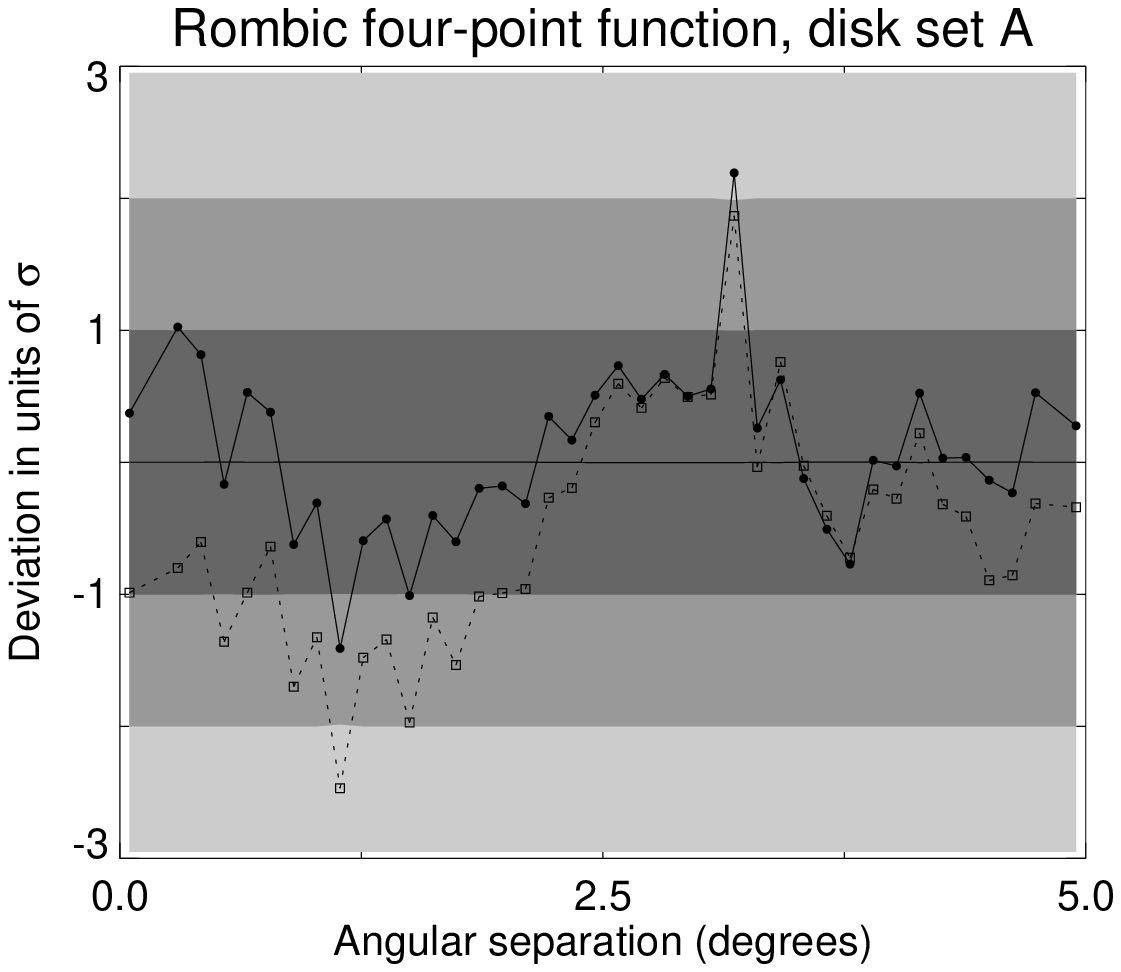,width=0.25\textwidth,clip=}}
\subfigure{\label{fig:selfunc_intermed_gauss_fourpt_diskB}\epsfig{figure=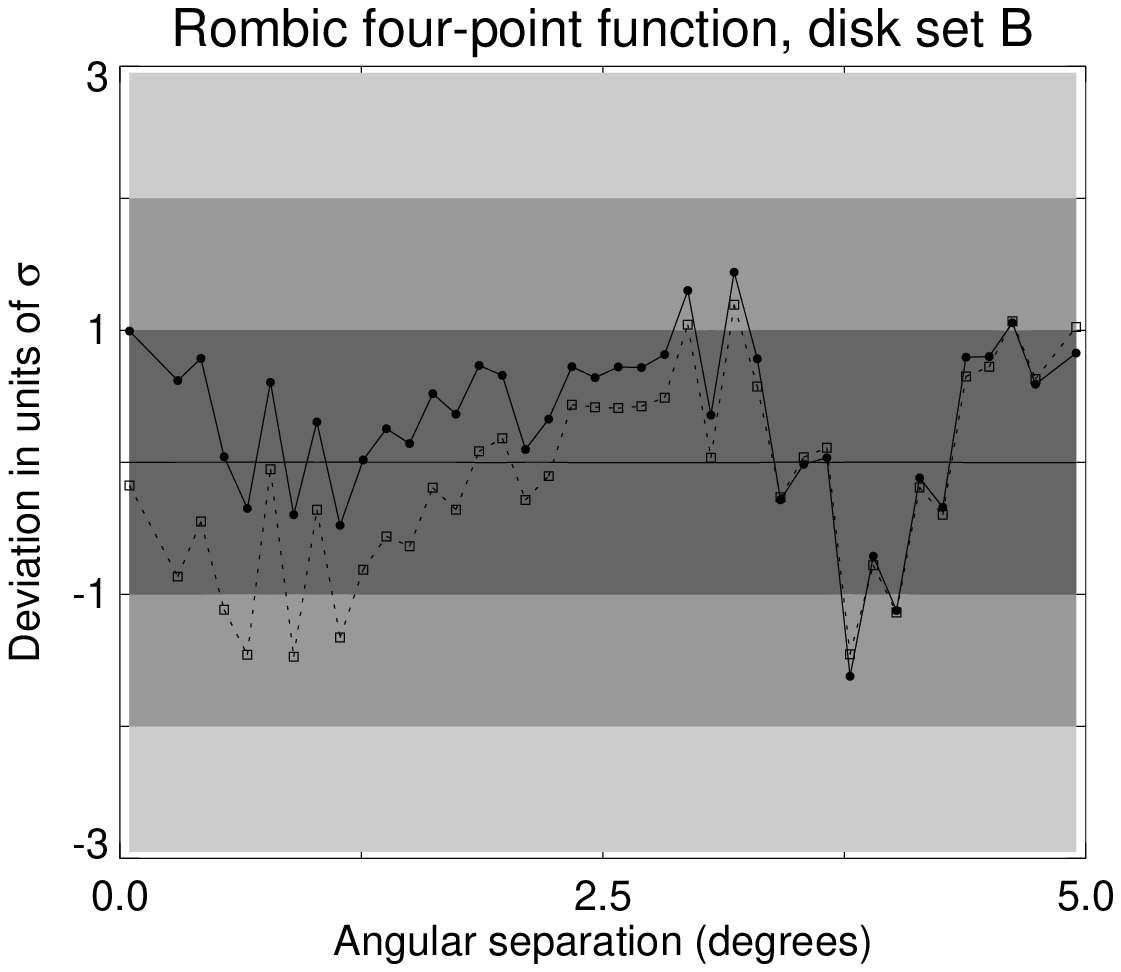,width=0.25\textwidth,clip=}}}

\caption{The full-sky intermediate-scale correlation functions. The
left-hand column shows the functions directly as measured from the
union of the two disk sets, while the normalized functions (see
equation \ref{eq:gaussianize2}) are shown in the other three
columns. The gray bands indicate 1, 2, and $3\,\sigma$ bands, as computed
from simulations. The dotted line corresponds to the foreground
corrected map and the solid line to the uncorrected map.}
\label{fig:selected_functions_intermed}
\end{figure*}

In Figure \ref{fig:selected_functions_intermed} the full-sky,
intermediate-scale two-point, equilateral three-point and rhombic
four-point functions are plotted, as computed from equation
\ref{eq:fullsky_avg}. The corresponding $\chi^2$ results are shown in
Table \ref{tab:chisq_intermed}. We see that the agreement between
observations and simulations is in all cases very good, both in terms
of overall shape and amount of small-scale fluctuations. Note,
however, that these figures only show a small subset of the
configurations included in the full analysis; while there are 41
three-point configurations with a base line of $5^{\circ}$, there are
only two with a $10'$ base line. Thus, the right hand sides of the
three- and four-point function plots are weighted much more strongly
than the left hand side in the $\chi^2$ analysis. Generally speaking,
visualizing higher-order functions is difficult because of their high
dimensionality.

The results from the corresponding $\chi^2$ analysis are shown in
Table \ref{tab:chisq_intermed}. Here we see that the results for the
foreground corrected map all lie comfortably between 0.05 and 0.95,
and no sign of discrepancy is found. For the raw maps, the $\chi^2$
numbers are generally somewhat high, but not disconcertingly so. The
agreement with the assumed model on intermediate scales appears to be
satisfactory on intermediate scales, as far as $N$-point correlation
functions are concerned.

\begin{figure*}
\mbox{\epsfig{file=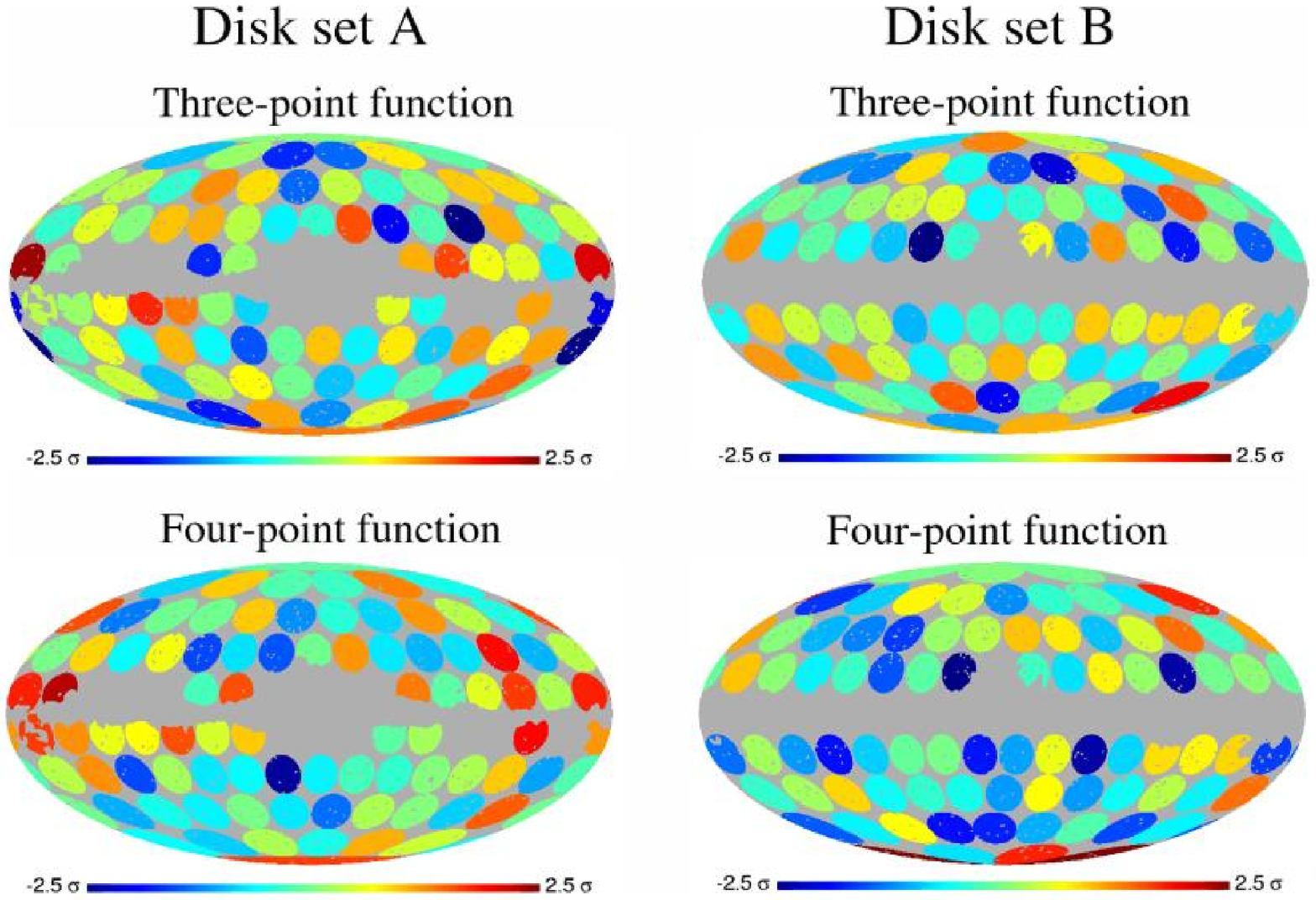,width=\linewidth,clip=}}
\caption{Results from the small-scale correlation function
  analysis. The elements have the same meaning as in figure
  \ref{fig:intermed_results}. Note the high $\chi^2$ values around the
  galactic plane in disk set A.}
\label{fig:smallscale_results}
\end{figure*}

A similar analysis, including disks in the galactic or ecliptic
hemispheres only, was also performed, but it did not find any clear
discrepancy in either case. Thus, the asymmetric pattern seen in
Figure \ref{fig:intermed_results} in the three-point function in set B
does not correspond directly to a dipole type distribution. Of course,
we could subdivide the sky further according to the observed patterns,
but this would strongly dilute the final probabilities, since we then
define our test \emph{a-posteriori}. All in all, the intermediate
scale correlation functions accepts the model, although hints of
hemisphere asymmetry may be seen by eye in the three-point function.

\subsection{Small-scale analysis}

The analysis from the previous section is now repeated, but this time
including \emph{all} three-point configurations with a longest edge
shorter than or equal to $72'$ (about 220 different configurations),
and twice as many four-point configurations. The four-point
configurations are defined in terms of the three-point configurations,
by letting the fourth point either be mirrored or rotated about the
base line, that once again is defined to be the longest edge of the
triangle.

\begin{figure*}
\center
\mbox{
\subfigure{\label{fig:hist_A_small_threept}\epsfig{figure=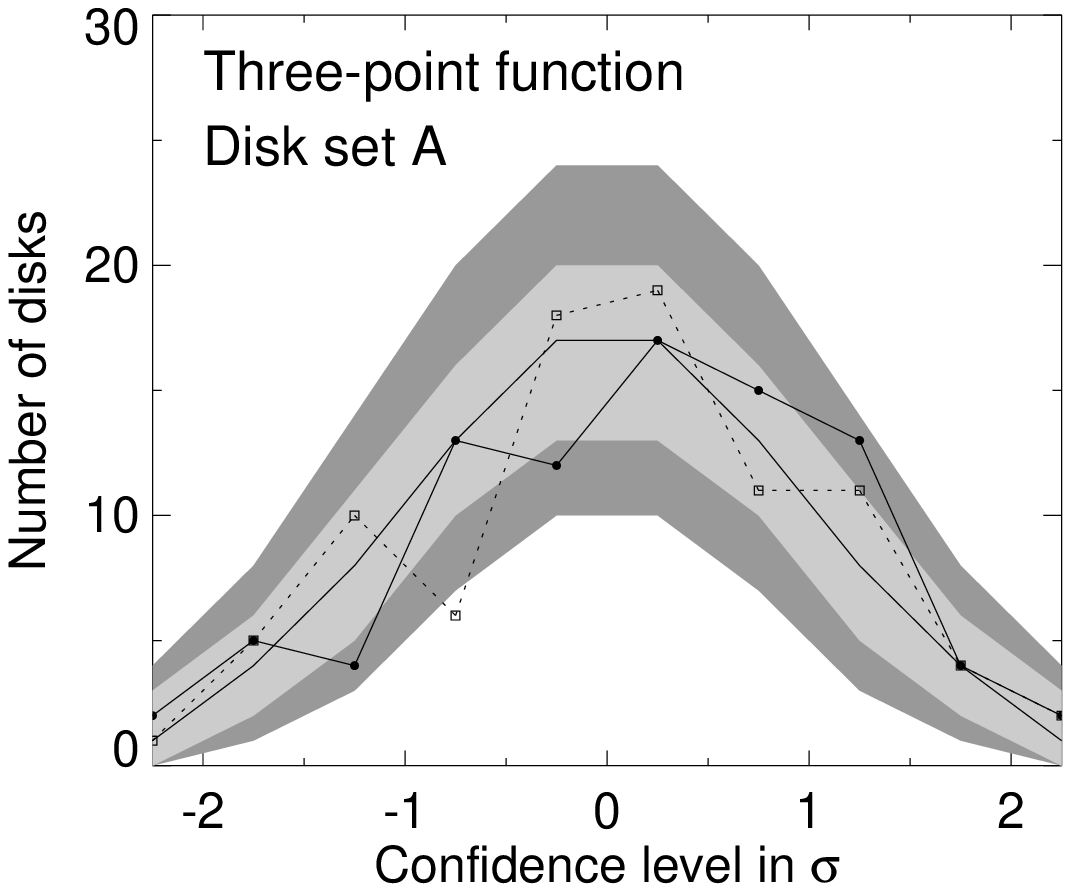,width=0.31\textwidth,clip=}}
\subfigure{\label{fig:hist_A_small_fourpt}\epsfig{figure=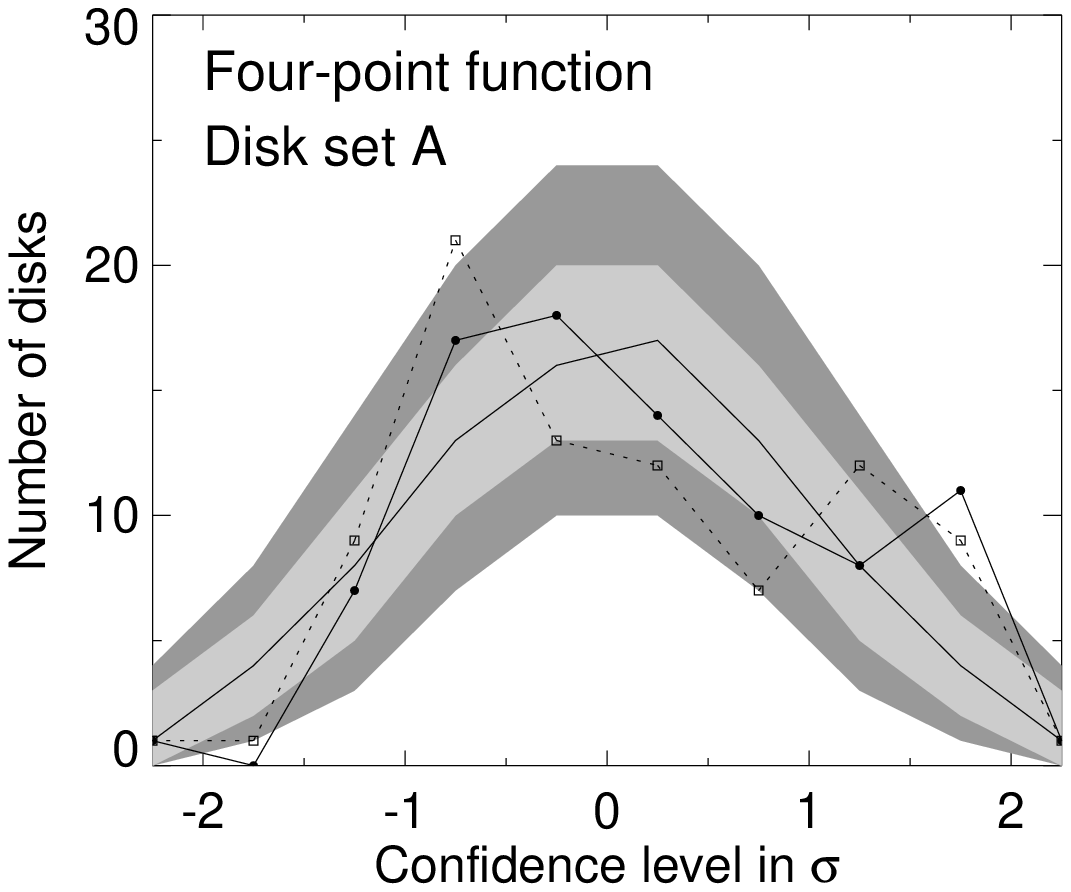,width=0.31\textwidth,clip=}}
}

\mbox{
\subfigure{\label{fig:hist_B_small_threept}\epsfig{figure=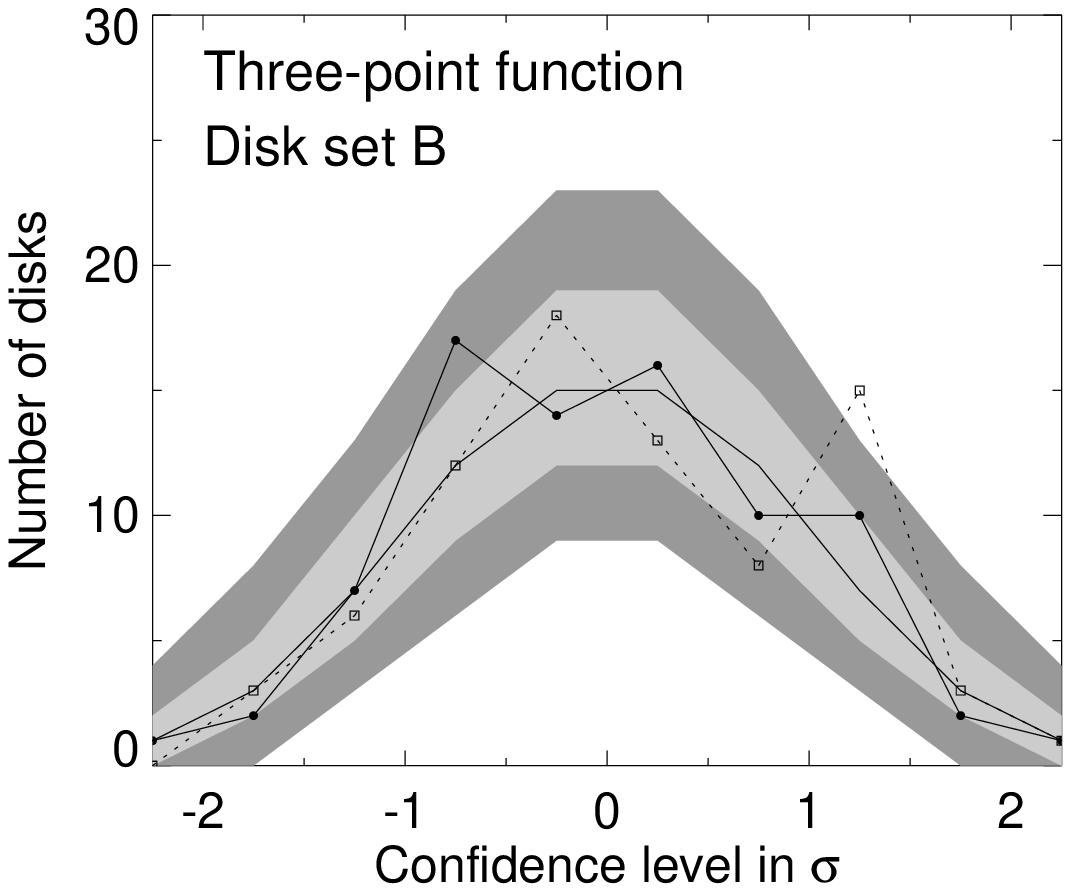,width=0.31\textwidth,clip=}}
\subfigure{\label{fig:hist_B_small_fourpt}\epsfig{figure=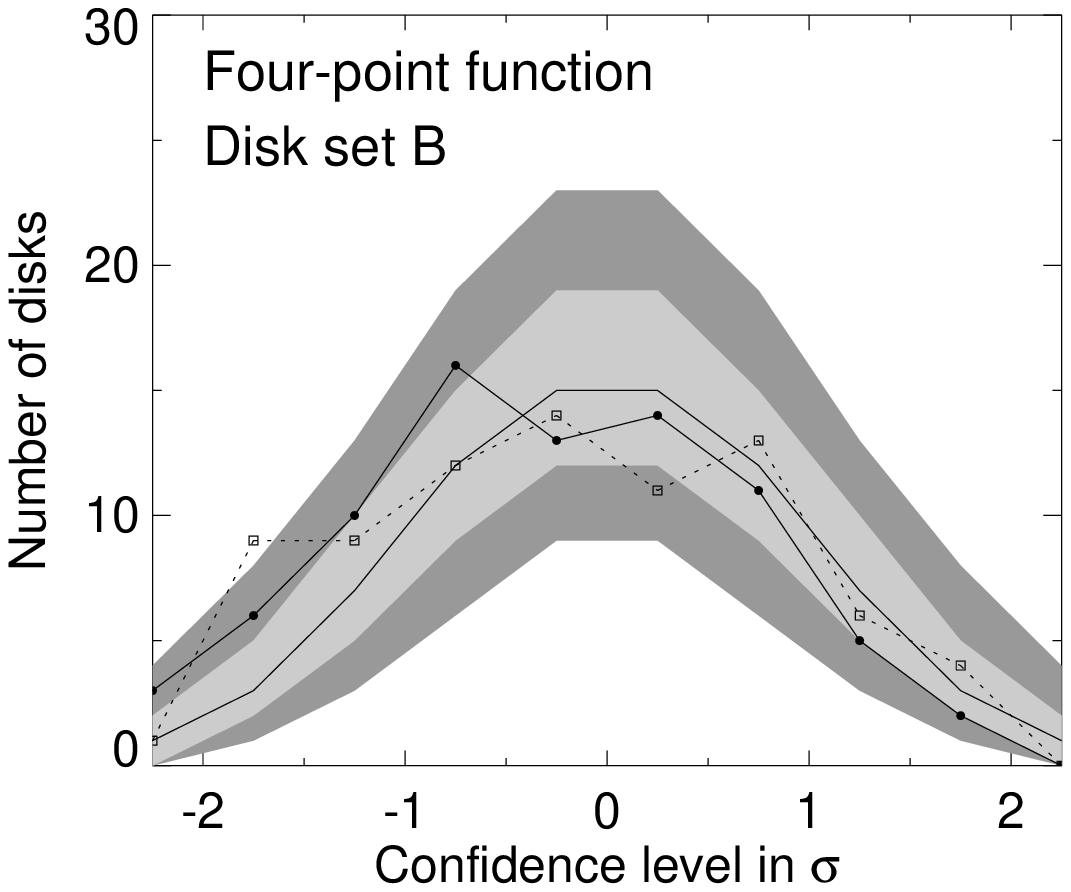,width=0.31\textwidth,clip=}}
}
\caption{Distributions (histograms) of the small scale disk confidence
  levels. The top row shows the results for disk set A, the bottom for
  disk set B. The columns show the three- and four-point function
  results, respectively. Gray bands indicate 1 and $2\,\sigma$
  confidence regions, computed from 5000 simulations, the solid lines
  indicate the results from the foreground corrected map, and the
  dotted lines show the results from the raw co-added map.}
\label{fig:hist_A_small}
\end{figure*}

The results from this disk-based analysis are shown in Figure
\ref{fig:smallscale_results}, and distributions of the corresponding
$\chi^2$ values are plotted in Figure \ref{fig:hist_A_small}. By eye,
the three-point function results in Figure
\ref{fig:smallscale_results} appear to be in quite good agreement with
the model, and no clear anomalies stand out. This impression is
confirmed both by the full-sky $\chi^2$ numbers, as well as by the
plots showing the disk $\chi^2$ distributions, except for the fact
that there are quite a large number of disks in the 1--$1.5\,\sigma$
range in disk set B.

However, the four-point function plot for disk set A shows a more
interesting effect; at least 7 out of the 21 disks touching the
galactic sky cut in disk set A have a fairly high $\chi^2$ value, and
there are no disks with low $\chi^2$ values. This is most likely an
indication of residual foregrounds near the galactic plane, a
conclusion which becomes even more plausible when studying figure 11
in the paper by \citet{bennett:2003b}. In these plots, clear residuals
are seen outside the Kp2 mask, particularly in the $Q$-band map.

\begin{deluxetable}{lccc}
\tablewidth{0pt}
\tabletypesize{\small}
\tablecaption{Intermediate scale $\chi^2$
results\label{tab:chisq_intermed}} 
\tablecomments{Results from the intermediate scale full-sky $\chi^2$
tests. The numbers indicate the fraction of simulated realizations
with $\chi^2$ value \emph{lower} than that for the co-added
\emph{WMAP} map. The upper half shows the results before correcting
for foregrounds, and the lower half shows the results after applying
foreground corrections.}
\tablecolumns{4}
\tablehead{Correlation function & Both sets & Disk set A & Disk set B  }
\startdata

\cutinhead{No foreground correction}
Two-point               & 0.829 & 0.322 & 0.944 \\
Three-point function 	& 0.955 & 0.805 & 0.981 \\
Four-point function	& 0.941 & 0.917 & 0.722 \\
\cutinhead{Foreground correction by external templates}
Two-point function      & 0.756 & 0.189 & 0.901 \\
Three-point function 	& 0.816 & 0.527 & 0.938 \\
Four-point function	& 0.683 & 0.674 & 0.330 

\enddata

\end{deluxetable}

\begin{deluxetable}{lccc}
\tablewidth{0pt}
\tabletypesize{\small}
\tablecaption{Small scale $\chi^2$
results\label{tab:chisq_small}} 
\tablecomments{Results from the small scale full-sky $\chi^2$
tests. The numbers indicate the fraction of simulated realizations
with $\chi^2$ value \emph{lower} than that for the co-added
\emph{WMAP} map. The upper half shows the results before correcting
for foregrounds, and the lower half shows the results after applying
foreground corrections.}
\tablecolumns{4}
\tablehead{ & Both sets & Disk set A & Disk set B  }
\startdata

\cutinhead{No foreground correction}
Three-point function 	& 0.725 & 0.600  & 0.554 \\
Four-point function	& 0.325 & 0.507 & 0.360 \\

\cutinhead{Foreground correction by external templates}
Three-point function 	& 0.330 & 0.084 & 0.214 \\
Four-point function	& 0.177 & 0.639 & 0.317 

\enddata

\end{deluxetable}

We may quantify the significance of this effect by computing a new
disk-averaged correlation function. This time we include only those 21
disks in the two near-galactic rows (A34--54), and the corresponding
results are shown in Table \ref{tab:chisq_galplane} for both
intermediate and small scales.

The four-point function results have a combined significance at
$2\,\sigma$ for the intermediate scales, and almost $3\,\sigma$ for the
small scales. In fact, for the small scales even the three-point
function has a $\chi^2$ value at the $2\,\sigma$ level. From these
considerations, it seems likely that the simple foreground correction
method by templates discussed by \citet{bennett:2003b} leaves
significant residuals near the galactic plane. Indeed, this should not
be surprising since the input synchrotron and free-free templates do
not contain power on the small angular scales probed by the $Q$-, $V$-,
and $W$-band maps, and the template fitting method itself does not admit
spectral variations on the sky, while it is likely that such
variations are seen close to the galactic plane.

We also compute the full-sky, disk-averaged correlation function for
the small-scale functions, and the results from this $\chi^2$ analysis
are shown in Table \ref{tab:chisq_small}. Here we see that the model is
comfortably accepted on these scales, and the effect of the foreground
residuals discussed above is diluted by the additional sky coverage.

\begin{deluxetable}{lccc}
\tablewidth{0pt}
\tabletypesize{\small}
\tablecaption{Galactic plane $\chi^2$ results\label{tab:chisq_galplane}} 
\tablecomments{Results from $\chi^2$ tests of the correlation functions
  computed over the disks near the galactic plane (disks A34-54).}
\tablecolumns{4}
\tablehead{Scales & Two-point & Three-point & Four-point}
\startdata

Intermediate    & 0.533   & 0.826 & 0.975 \\ 
Small           & \nodata & 0.975 & 0.998 

\enddata

\end{deluxetable}

Finally, we make one connection to a previously reported detection of
non-Gaussianity \citep{vielva:2004,cruz:2004}: A very cold spot was
found at galactic coordinates $(b=-57^{\circ}, l=207^{\circ})$ using
wavelet statistics, this corresponds to disk number B73 (see Figure
\ref{fig:intermed_results}) in our partitioning of the sky. This
particular disk has a three-point function $\chi^2$ that is high at
the $2\,\sigma$ level on both intermediate and small scales,
insignificant by itself, yet perhaps interesting when taken in
combination with the \citet{vielva:2004} detection.

\section{Conclusions}
\label{sec:conclusions}

We have computed the two-, three-, and four-point correlation
functions from the first-year \emph{WMAP} data sets, and find
interesting effects on several angular scales.  On the very largest
scales an asymmetric distribution of power is observed in both the
two-, three- and four-point functions, in that the fluctuations on the
southern ecliptic (and galactic) hemisphere are significantly stronger
than on the northern ecliptic (and galactic) hemisphere. In order to
study this effect more closely, we computed the correlation functions
from each frequency band separately, and for two different sky cuts,
and found that the asymmetry is present in any of the bands, and
independent of the particular region definition. This argues against a
foreground based explanation for this effect.

Next, we computed the two-point correlation functions from a set of
difference maps, and detected excess correlations in the data among
the $W$-band differencing assemblies, which are not accounted for in the
detailed simulation pipeline used by the \emph{WMAP} team. While this
effect could potentially pose a serious problem for the upcoming
polarization data, its absolute amplitude is very small compared to
the temperature anisotropy amplitude, and it is therefore highly
unlikely to cause any problems for results based on the first-year
\emph{WMAP} temperature data.

We then computed the correlation functions on small ($<72\arcmin$) and
intermediate ($<5^{\circ}$) scales, and found that the agreement with
the Gaussian model is generally good in these cases. Although a
pattern consistent with the large-scale asymmetry discussed earlier is
visible in the intermediate-scale three-point correlation function, it
is difficult to assess the significance of this pattern. It should be
regarded more as supportive evidence to the large-scale results, than
as a conclusive result on its own. Overall, the Gaussian model is
accepted by the intermediate scale $N$-point correlation functions.

On small scales we detect residual foregrounds near the galactic plane
roughly at the $2.5\sigma$ level. However, such residuals may be seen
by eye in the actual maps, and this is therefore not a surprising
result. Except for this residual foreground detection the model is
accepted by $N$-point correlation functions on the smallest scales
probed in this paper.

As seen from the analyses presented in this paper, real-space based
statistics, such as the $N$-point correlation function, have a clear
value with respect to control of systematics. For cosmological
purposes, harmonic-space statistics (e.g., the angular power spectrum,
and the bi-spectrum) are usually the preferred tools since they
generally have a simpler physical interpretation than their real-space
counterparts. However, systematics are often localized in real space
rather than in harmonic space (e.g., foregrounds are highly localized
in space; $1/f$ noise leads to stripes along the scan directions;
cross-talk between detectors leads to noise correlations at some given
scale), and real-space statistics can therefore often be more powerful
for detecting their presence. The results presented in this paper are
clear demonstrations of this fact.

\begin{acknowledgements}
The authors thank Gary Hinshaw and Pablo Fosalba for useful
discussions. H.~K.~E.\ thanks Dr.\ Charles R.\ Lawrence for much
support, and especially for arranging his visit to JPL. He also thanks
the Center for Long Wavelength Astrophysics at JPL for its hospitality
while this work was completed. The authors acknowledge use of the
HEALPix software and analysis package for deriving the results in this
paper, and use of the Legacy Archive for Microwave Background Data
Analysis (LAMBDA).  H.~K.~E.\ and P.~B.~L.\ acknowledge financial
support from the Research Council of Norway, including a Ph.~D.\
studentship for H.~K.~E.  This work has received support from The
Research Council of Norway (Programme for Supercomputing) through a
grant of computing time.  This work was partially performed at the Jet
Propulsion Laboratory, California Institute of Technology, under a
contract with the National Aeronautics and Space Administration. This
research used resources of the National Energy Research Scientific
Computing Center, which is supported by the Office of Science of the
U.S. Department of Energy under Contract No. DE-AC03-76SF00098.
\end{acknowledgements}

\end{document}